\documentclass[a4paper,UKenglish,cleveref, autoref, thm-restate]{lipics-v2021}

\usepackage{subcaption}
\usepackage{multirow}
\usepackage{tikz}
\usepackage{xspace}
\usepackage{enumitem}
\usepackage{tabularray}

\newcommand{\ltsat}{\textsc{Linear 3-SAT}\xspace}
\newcommand{\lsat}{\textsc{LSAT}\xspace}
\newcommand{\no}[1]{\ensuremath{\overline{#1}}}

\tikzstyle{noeud}=[circle,inner sep=1, minimum size =20 pt, line width = 1pt, draw=black, fill=white]
\tikzstyle{exit}=[circle,inner sep=2, minimum size =3 pt, line width = 1pt, draw=black, fill=red]

\title{Nemesis, an Escape Game in Graphs} 

\author{Pierre Berg\'e}{Univ. Grenoble Alpes, CNRS, Grenoble INP, LIG, 38000 Grenoble, France\and \url{https://lig-membres.imag.fr/bergep/} }{Pierre.Berge@univ-grenoble-alpes.fr}{https://orcid.org/0000-0002-6170-9473}{}

\author{Antoine Dailly}{Université Clermont Auvergne, INRAE, UR TSCF, 63000, Clermont-Ferrand, France\and \url{https://daillya.github.io/}}{antoine.dailly@inrae.fr}{}{}

\author{Yan Gerard\footnote{Corresponding author}}
 	{Université Clermont-Auvergne, CNRS, Mines de Saint-Étienne, Clermont-Auvergne-INP, LIMOS, 63000 Clermont-Ferrand, France}
 	{yan.gerard@uca.fr}
 	{https://orcid.org/0000-0002-2664-0650}
 	{}

\authorrunning{P. Berg\'e, A. Dailly, Y. Gerard} 

\Copyright{Pierre Berg\'e, Antoine Dailly, Yan Gerard} 



\category{} 

\relatedversion{} 

\acknowledgements{We  thank Claire Hilaire and Gaëtan Berthe for the discussions and the different ideas they brought to this work. }

\nolinenumbers 

\EventEditors{John Q. Open and Joan R. Access}
\EventNoEds{2}
\EventLongTitle{}
\EventShortTitle{}
\EventAcronym{}
\EventYear{}
\EventDate{}
\EventLocation{}
\EventLogo{}
\SeriesVolume{X}
\ArticleNo{N}

\begin{document}

\maketitle

\begin{abstract}
We define a new escape game in graphs that we call \textit{\textsc{Nemesis}}. The game is played on a graph having a subset of vertices labeled as exits and the goal of one of the two players, called the \textit{fugitive}, is to reach one of these exit vertices. The second player, \emph{i.e.} the fugitive adversary, is called \textit{the Nemesis}. Her goal is to trap the fugitive in a connected component which does not contain any exit. At each round of the game, the fugitive moves from one vertex to an adjacent vertex. Then the Nemesis deletes one edge anywhere in the graph. The game ends when either the fugitive reached an exit or when he is in a connected component that does not contain any exit.  
In trees and graphs of maximum degree bounded by $3$, \textsc{Nemesis} can be solved in linear time. We also show that a variant of the game  called \textsc{Blizzard} where only edges adjacent to the position of the fugitive can be deleted also admits a linear time solution. For arbitrary graphs, we show that \textsc{Nemesis} is PSPACE-complete, and that it is NP-hard on planar multigraphs. We extend our results to the related \textsc{Cat Herding} problem, proving its PSPACE-completeness. We also prove that finding a strategy based on a full binary escape tree whose leaves are exists is NP-complete.
\end{abstract}

\section{Introduction}
\label{sec:typesetting-summary}

\paragraph*{Nemesis escape game}

Over the last twenty years, Escape Games have become a highly popular form of entertainment in many countries. These are collaborative games in which a participant or a team attempts to exit a room by solving a sequence of riddles. In this paper, we propose a similar framework, but set in a graph environment. An individual, whom we call the \textit{fugitive}, seeks to escape from a graph. Some vertices are labeled as \textit{exits}, and the objective of the fugitive is to reach any of them.

At each round of the game, the fugitive moves from his current vertex to an adjacent one by traversing an edge. If, at any round, he manages to reach an exit, he wins. In a static graph, a simple breadth-first search (BFS) computes the shortest path from the fugitive’s starting vertex to an exit, giving the exact minimum number of rounds required to escape.

\begin{figure}[ht]
  \centering
  \begin{subfigure}[t]{\textwidth}
		\includegraphics[width=\textwidth]{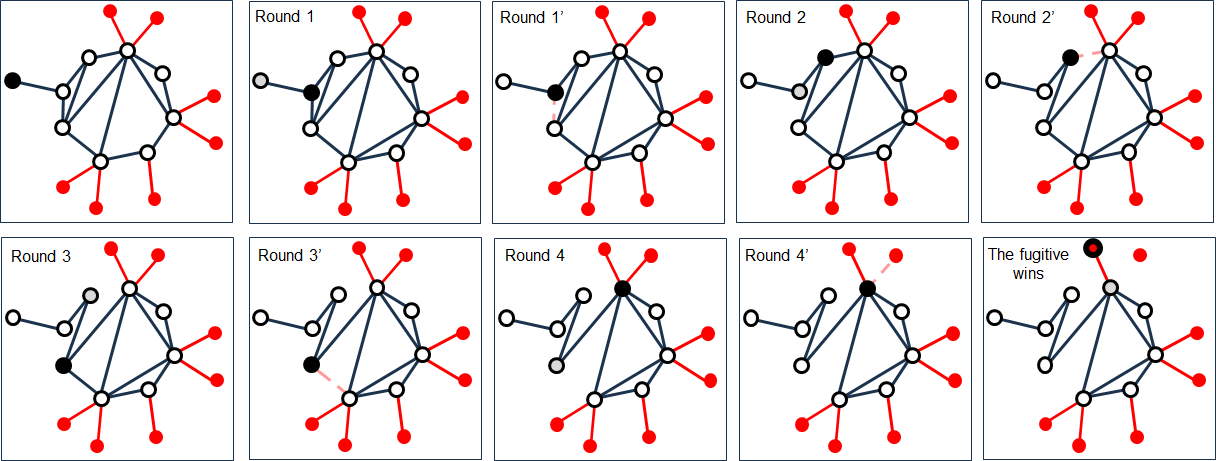}
        \caption{In the above game, the fugitive wins.}
  \end{subfigure}
  ~
    \begin{subfigure}[t]{\textwidth}
    \includegraphics[width=\textwidth]{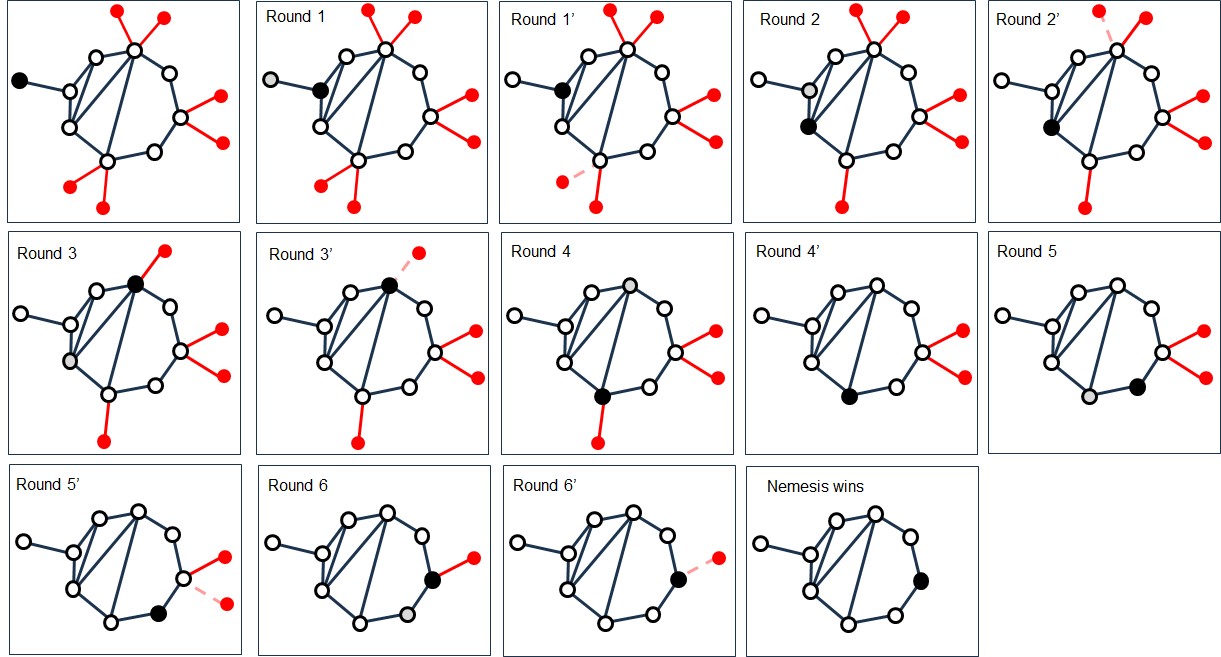}
    \caption{In the above game, the Nemesis wins.}
    \end{subfigure}
	\caption{\label{examples} \textbf{Two games of \textsc{Nemesis}.} Exits are red vertices. At each round, the fugitive whose position is represented by the black vertex moves from one edge while the Nemesis removes one edge.}
\end{figure}

The  game comes from the introduction of an adversary of the fugitive who has the magic power to alter the graph. We now define the new two player's  escape game in a graph that we call \textit{Nemesis}\footnote{In Greek mythology, Nemesis is a goddess of righteous anger and divine punishment.}. \textit{The Nemesis}\footnote{We add the article \textit{the} into \textit{the Nemesis} for referring to the fugitive adversary while \textit{\textsc{Nemesis}} without article is used as name of the game} is the name of the adversary of the fugitive. The Nemesis alters the structure of the graph in a very restricted and natural way. At each round, after the fugitive's move, the Nemesis permanently removes exactly one edge of the graph. The deleted edge can be chosen anywhere in the graph. The Nemesis wins if she  succeeds in cutting all paths from the fugitive’s position to every exit, thereby confining him in the graph forever. The game ends either when the fugitive reaches an exit or when the Nemesis traps him in a connected component without any exit (two plays are represented in  Fig.~\ref{examples}).

Despite the simplicity of this setting, and although many pursuit-and-evasion or blocking games on graphs have been studied, \textsc{Nemesis} is a natural escape game which, to the best of our knowledge, has never been investigated before. However, \textsc{Nemesis} is closely related to several well-studied problems.

\paragraph*{Related Games}

There is a rich literature on pursuit games on graphs, in particular on the classical \textsc{cops and robber problem}~\cite{quillot1978jeux,aigner1984game,meyniel1987graphes,fomin2010cop,bonato2011game,bowler2021bounding}, which has been proved to be PSPACE-complete in~\cite{mamino2012computationalcomplexitygamecops}. There are two major differences between \textsc{cops and robber} games and \textsc{Nemesis}. First, in the \textsc{cops and robber problem}, there are no exit vertices. Second, the pursuers move on the graph, whereas in our setting the Nemesis instead deletes edges.

\textsc{Nemesis} is more closely related to the \textsc{guard problem}, which is also PSPACE-complete~\cite{Fom08,Fom11}. In the guard problem, the input is a graph  with a subset of vertices defining a protected area, defended by a fixed number of guards. The goal of the guards is to prevent an intruder from reaching this protected area. If one replaces the intruder by a fugitive and the protected area by exits, the framework is very similar to \textsc{Nemesis}. However, the essential difference is that the guards move on the vertices of the graph, while in our simple game the Nemesis irrevocably deletes edges.

Another related problem is John Conway’s \textsc{Angel and Devil problem}, in which an angel moves on an infinite two-dimensional grid while the devil attempts to trap it by progressively removing (or “eating”) vertices~\cite{Angel1,Angel2}. While it has been shown that the angel wins if they can move at distance at least~2 each round~\cite{Angel3,mathe2007angel,wastlund2008weaker}, several variants have been studied depending on the respective powers of the angel and the devil~\cite{Angel4} or in other graphs such as 3D grids~\cite{kutz2005conway,bollobas2006angel}. In the same spirit of trapping a player moving in a graph, the \textsc{Cat Herding} problem was recently introduced in~\cite{cats} and investigated in~\cite{cat2,cattree}. Instead of catching an angel in an infinite grid, the herder tries to trap a cat maliciously wandering in a finite graph. This problem looks like \textsc{Nemesis} in that the herder deletes edges one at a time while the cat moves from one edge at each round. The significant difference is that the cat has no exit to escape to, which could be unsurprising for a kitten, but unfortunate for a rooftop cat. The absence of exit vertices makes \textsc{Cat Herding} fundamentally different from \textsc{Nemesis}.  In the \textsc{Cat Herding} problem, the goal of the cat is to remain free for as long as possible, while the herder attempts to trap it as quickly as possible. A by-product of the results about \textsc{Nemesis} is the PSPACE-completeness of the \textsc{Cat Herding} problem. We also note the existence of the 2012 commercial board game \textsc{Nowhere to Go}~\cite{nowhereToGo}, in which two players try to trap their opponent by removing edges in a finite hexagonal grid.

\textsc{Nemesis} may also be viewed as a connectivity game, close to \textsc{Hex} or \textsc{Generalized Geography}, since one player (the fugitive) attempts to build a connection between its initial vertex and an exit while avoiding the edges deleted by the Nemesis. The difference is that the fugitive must follow a single path, restricting his choices more than in unstructured vertex-selection games. Many connectivity games are known to be PSPACE-complete, such as \textsc{Hex} on general~\cite{Hex1} and on planar graphs~\cite{Hex2}, or \textsc{Havannah}, \textsc{Twixt}, and \textsc{Slither} \cite{Bon16}.

A final closely related problem is the \textsc{Canadian Traveler Problem}, introduced by Papadimitriou and Yannakakis in 1991~\cite{Papa}. In this online problem, a traveler wishes to go from its initial position $s$ to one target vertex $t$ on a known graph, but each edge may be blocked by snow, and the traveler only learns whether an edge is passable upon reaching one of its endpoints. 
The task of designing a good strategy can be modeled as a game between the traveler and the storm. The traveler moves in the graph exactly as the fugitive in our escape game, while the snowstorm plays the role of the adversary who determines the availability of edges adjacent to the traveler’s current vertex. The traveler aims to keep the total length of his walk within a given competitive ratio compared to the optimal available path. Papadimitriou and Yannakakis proved that the \textsc{Canadian Traveler Problem} is PSPACE-complete~\cite{Papa}. The first difference with \textsc{Nemesis} is in the objective of the Canadian traveler which is to maintain an effective competitive ratio when it reaches the exit. In contrast, the goal of the fugitive is simply to reach to an exit: the notion of competitive ratio is thus irrelevant in \textsc{Nemesis}. Another difference is in the power of the adversary: while the Nemesis can remove an edge anywhere on the graph, the storm only impacts edges incident with the traveler's position when he first reaches it. A last difference is that the snow can remove many edges in one round while the Nemesis can only remove them one by one. 


As nod of the \textsc{Canadian Traveler Problem}, we introduce a variant of \textsc{Nemesis} that we call \textsc{Blizzard}. The Nemesis now takes the form of a blizzard and the fugitive is a trapper searching for a shelter. This variant mirrors the Canadian Traveler setting: the trapper tries to escape to a shelter, while at each round, the blizzard deletes an edge which is now incident with the fugitive’s current vertex (in \textsc{Nemesis}, any edge can be deleted while in \textsc{Blizzard}, we have this condition that only an edge which is incident with the trapper can be removed).
The blizzard wins if the trapper is doomed to freeze to death in the snow, because all his paths to shelter have been made impassable.
We show that this restriction on the power of the Nemesis makes \textsc{Blizzard} significantly more tractable.

\paragraph*{Results}

\textsc{Nemesis} and \textsc{Blizzard} are zero-sum games.
Our main question is to determine, for a given starting position of the fugitive (respectively the trapper), which player has a winning strategy.
By solving the game, we mean deciding which of the two players can force a win from the given starting vertex.

An overview of our results on \textsc{Nemesis} (and \textsc{Blizzard}) is given in \Cref{tab:results}. They fall into two classes: polynomial-time solvable cases, and hardness results obtained by reductions from computationally hard problems.

A key concept throughout the paper is that of a \emph{binary escape tree}.
A binary escape tree is a full binary tree in which the root has degree~$2$, all internal vertices have degree~$3$, and all leaves are exits.
The presence of a binary escape tree in the neighborhood of the fugitive’s starting position guarantees a winning strategy for the fugitive.
However, this condition is not universal: there exist instances in which the fugitive has a winning strategy despite the absence of any nearby binary escape tree.

We first show that, for trees and for graphs of maximum degree~$3$ (without multiedges), the existence of a binary escape tree is both necessary and sufficient to certify that the fugitive wins.
Moreover, binary escape trees can be detected in linear time in both graph classes.
As a consequence, \textsc{Nemesis} can be solved efficiently on these graphs.

\begin{theorem}\label{t1}
For trees and graphs of maximum degree at most~$3$, \textsc{Nemesis} can be solved in linear time.
\end{theorem}

In the same spirit, for arbitrary graphs, \textsc{Blizzard} can also be solved in linear time using a simple graph traversal algorithm.

\begin{theorem}\label{t2}
\textsc{Blizzard} can be solved in linear time.
\end{theorem}

We then turn to hardness results and establish three negative results.
First, we show that the strategy consisting in searching for a binary escape tree (which is sufficient, but not necessary, to certify a winning strategy for the fugitive) is itself computationally hard.

\begin{restatable}{theorem}{binaryexit}
Given a graph with a set of exits and a vertex $r$, determining whether $r$ is the root of a binary escape tree is NP-complete, even when the graph has maximum degree~$4$.
\label{antoine}
\end{restatable}

We now present the main hardness contributions of the paper, which consist of two complexity results for \textsc{Nemesis}.
We first show that the game becomes computationally hard on planar multigraphs.

\begin{theorem}\label{t3}
For planar multigraphs, \textsc{Nemesis} is NP-hard.
\end{theorem}

However, as is often the case for two-player games, \textsc{Nemesis} is not known to belong to NP, even when restricted to planar multigraphs.
Therefore, NP-completeness is not the appropriate complexity notion for this problem.
In fact, the exact complexity of \textsc{Nemesis} on planar graphs (even without multiedges) remains open and appears to be a challenging question.

The proof of Theorem~\ref{t3} relies on a reduction from a planar variant of \textsc{SAT}.
Unfortunately, we are currently unable to reduce the seminal PSPACE-complete problem \textsc{Quantified SAT} (\textsc{QSAT}) to \textsc{Nemesis} in planar multigraphs, and thus cannot establish PSPACE-completeness in this restricted setting.

Nevertheless, we provide a reduction from \textsc{QSAT} to \textsc{Nemesis} in arbitrary multigraphs.
The additional gadget required to simulate quantifier alternation is inherently non-planar.
As a consequence, we obtain PSPACE-completeness only for arbitrary multigraphs and, after additional arguments, for arbitrary simple graphs.

\begin{theorem}\label{t4}
\textsc{Nemesis} is PSPACE-complete.
\end{theorem}

\begin{table}[t]
    \centering
    
    
    
    \begin{tblr}{
		colspec={Q[c,4em]Q[c,5em]Q[c,5em]Q[c,7em]Q[c,10em]},
		rows=m,
		vlines,
		hlines,
		hline{2}={1pt},
		vline{2}={1pt}
	}
	
    	& \textbf{Trees} & $\mathbf{\Delta \leq 3}$ & \textbf{Planar graphs} & \textbf{Arbitrary graphs} \\
    	\textbf{Simple graphs} & \textcolor{green}{Linear} (Th.~\ref{t1}) & \textcolor{green}{Linear} (Th.~\ref{t1}) & ? & \textcolor{red}{PSPACE-complete} (Th.~\ref{t4} + Th.~\ref{twoexits}) \\
    	\textbf{Multi-graphs} & ? & ?  & \textcolor{red}{NP-hard} (Th.~\ref{t3}) & \textcolor{red}{PSPACE-complete} (Th.~\ref{t4})
    	
    \end{tblr}
    
    \caption{A summary of the complexity results for \textsc{Nemesis}.}
    \label{tab:results}
\end{table}

As a corollary of Theorem~\ref{t4}, we obtain the complexity of the \textsc{Cat Herding} problem.

\begin{corollary}
The \textsc{Cat Herding} problem is PSPACE-complete.
\end{corollary}

The paper is organized as follows.
In Section~\ref{sec:prelim}, we formally define the games, and give examples and basic observations.
Section~\ref{sec:poly} is devoted to graph classes for which \textsc{Nemesis} can be solved in polynomial time.
Finally, Section~\ref{sec:hard} contains our hardness results.

\section{Problem Statement and First Observations} \label{sec:prelim}

We begin with a formal definition of \textsc{Nemesis}. We also present preliminary remarks on straightforward cases or dealing with the number of exits.

\subsection{The Game}

We define the game \textsc{Nemesis} on either a graph $G=(V,E)$ or a multigraph (where $E$ is a multiset of edges).
A subset of vertices is designated as \textit{exits}, while the remaining vertices are called \textit{regular}.

\textsc{Nemesis} is a two-player, zero-sum game.
Player~1 is the \emph{fugitive}, and Player~2 is the \emph{Nemesis}.
The game is asymmetric: the fugitive moves along the graph, whereas the Nemesis removes edges.
Their objectives are also different.
The fugitive wins if he reaches an exit vertex, while the Nemesis wins if she isolates the fugitive in a connected component that contains no exit.

An instance of the game consists of a graph or multigraph $G$, a designated set of exits, and a starting vertex $s$ indicating the initial position of the fugitive.
We may assume without loss of generality that the exits form an independent set, since edges between exits play no role in the game.

The game proceeds in rounds.
At each round, the fugitive first moves from his current vertex $x$ to an adjacent vertex $y$ by traversing an edge $(x,y)$ still present in $E$.
Then, the Nemesis removes one edge from $E$.
In the case of multigraphs, if an edge $e$ has multiplicity $r \ge 1$, removing a copy of $e$ decreases its multiplicity to $r-1$.
If $r-1=0$, the edge disappears from the graph. Otherwise, it remains usable by the fugitive.
In the language of games, edges can be seen as having a number of \emph{health points} (HP).

Several variants of the game can naturally be defined.
For instance, the Nemesis could remove an edge before the fugitive makes the first move, or both players could act simultaneously without knowing the adversary's action.
Regardless of the precise timing rules, these variants are closely related.
Throughout the paper, we adopt the following convention: the fugitive moves first, thereby triggering the wrath of the gods and setting the Nemesis into action.

Since one edge is removed at each round, the number of rounds is bounded by the number of edges, and the game is finite.
\textsc{Nemesis} satisfies the following two basic properties.

\begin{remark}\label{noloop}
If the fugitive has a winning strategy, then he has a winning strategy that never visits the same vertex twice.
Indeed, suppose a winning strategy contains a loop between rounds $i$ and $j$.
Since fewer edges remain after round $j$ than after round $i$, the fugitive can skip the loop without weakening his position. In other words, adding the rule that the fugitive cannot come back to a previously visited vertex does not change the outcome of the game. This additional rule is taken into account in the proofs.
It follows that if the fugitive wins, he can do so in at most $|V|$ rounds.
As a consequence, in multigraphs, any edge of multiplicity greater than $|V|$ cannot be fully removed by the Nemesis.
Such edges behave as if they had infinite multiplicity and are called \emph{unbreakable}.
\end{remark}

\begin{remark}\label{no2exits}
In simple graphs, many instances are trivial.
If no vertex is adjacent to two exits, then the fugitive always loses, except in the trivial cases where he starts at an exit or at a neighbor of an exit.
Indeed, whenever the fugitive reaches a vertex at distance~$1$ from an exit, the Nemesis can remove the unique edge leading to the exit, making it unreachable.
Therefore, in simple graphs, non-trivial instances of \textsc{Nemesis} must contain vertices adjacent to at least two exits.
\end{remark}

\subsection{Exercises}

\begin{figure}[ht]
  \centering
  \includegraphics[width=\textwidth]{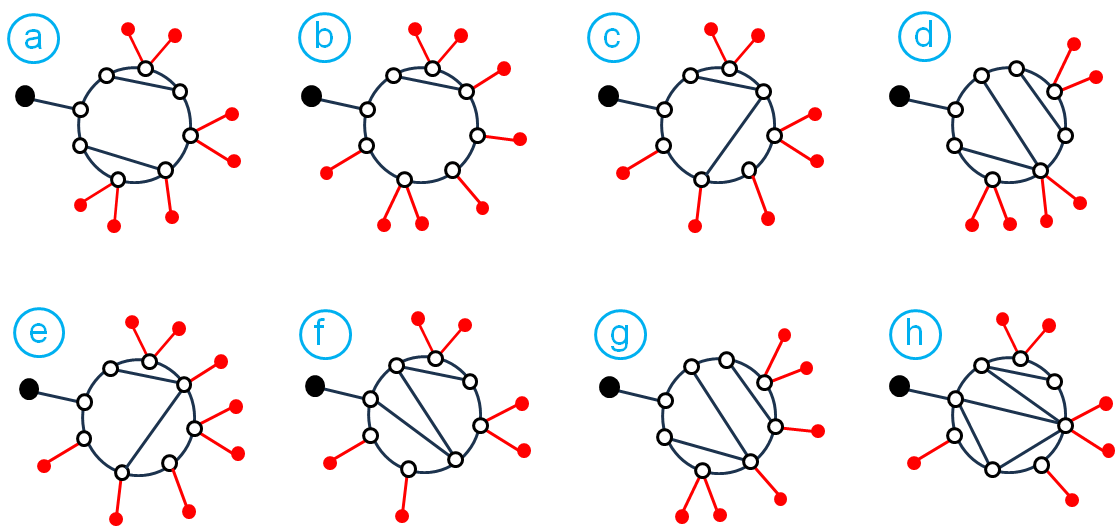}
  \caption{\textbf{Eight instances of \textsc{Nemesis}.} Can the fugitive, starting from the black vertex, escape?}
  \label{exercises}
\end{figure}

As a warm-up, we present several small instances of the game.
Who wins in the instances depicted in Fig.~\ref{exercises}?
Solutions are provided in the footnote.\footnote{a--Nemesis wins (N), b--Fugitive wins (F), c--N, d--N, e--F, f--N, g--F, h--F.}

\subsection{About the Number of Exits}

We now investigate how the number of exits influences the complexity of \textsc{Nemesis}.
As a first observation, there is a straightforward reduction from \textsc{Nemesis} to \textsc{Nemesis with one exit} in the case of multigraphs.
Indeed, all exits can be merged into a single exit vertex $t$ by replacing, for every vertex $v$, the edges from $v$ to exits by a single edge from $v$ to $t$ whose multiplicity equals the sum of the original multiplicities.

In a later section, we prove that \textsc{Nemesis} is PSPACE-complete on multigraphs (Theorem~\ref{t4}).
The reduction above immediately implies that even \textsc{Nemesis with one exit} is PSPACE-complete in this setting.
Consequently, assuming $\mathrm{P} \neq \mathrm{PSPACE}$, there cannot exist any tractable algorithm (neither FPT nor XP) parameterized by the number of exits.

\medskip

The situation is different for simple graphs.
With only one exit, \textsc{Nemesis} becomes trivial.
Indeed, there are two cases.
If the starting vertex $s$ is adjacent to the unique exit $t$, the fugitive wins immediately.
Otherwise, as noticed in Remark ~\ref{no2exits}, the Nemesis has a simple winning strategy: whenever the fugitive reaches a neighbor $u$ of $t$, she removes the edge $(u,t)$.

It is therefore natural to ask whether this observation extends to a fixed number of exits.
As we now show, the answer is negative: even in simple graphs, restricting the number of exits to two already captures the full complexity of the game.

\begin{theorem}\label{twoexits}
    For simple graphs, there is a polynomial-time reduction from \emph{\textsc{Nemesis}} to \emph{\textsc{Nemesis with two exits}}.
\end{theorem}

\begin{proof}
    Let $(G,s,X)$ be an instance of \textsc{Nemesis}, where $G=(V,E)$ is a simple graph, $s \in V$ is the starting vertex, and $X \subsetneq V$ is the set of exits, with $s \notin X$.
    We construct an instance $(G',s,\{t_1,t_2\})$ of \textsc{Nemesis with two exits} as follows.

    \begin{itemize}
        \item The vertex set of $G'$ is
        $
        V' = V \cup Y \cup \{t_1,t_2\},
        $
        where $Y=\{y_1,\dots,y_{n+1}\}$ is a set of $n+1$ new vertices.
        \item The edge set of $G'$ is
        $
        E' = E \cup (X \times Y) \cup (Y \times \{t_1,t_2\}),
        $
        that is, every vertex of $X$ is connected to every vertex of $Y$, and every vertex of $Y$ is connected to both exits $t_1$ and $t_2$.
        \item The starting position remains $s$, and the exits are $t_1$ and $t_2$.
    \end{itemize}

    The graph $G'$ has $2n+3$ vertices and at most $m + (n-1)(n+1) + 2(n+1)$ edges.

    We now show that the two instances are equivalent.
    First, assume that the fugitive has a winning strategy in $(G,s,X)$.
    By Remark~\ref{noloop}, he reaches some exit $x \in X$ within at most $n$ rounds.
    Up to this point, the Nemesis has removed at most $n$ edges.
    Since $|Y|=n+1$, there exists a vertex $y \in Y$ such that no edge incident to $y$ has been removed.
    In particular, $y$ is still adjacent to $x$, $t_1$, and $t_2$.
    The fugitive can therefore move from $x$ to $y$.
    As $y$ has two exit neighbors, the Nemesis can block at most one of them, and the fugitive wins.

	Conversely, if the Nemesis has a winning strategy in $(G,s,X)$, the same strategy applied in $(G',s,\{t_1,t_2\})$ allows her to prevent the fugitive to reach the vertices of $X$ and thus to reach any of the two exits $\{t_1,t_2\}$. 
\end{proof}

Theorem~\ref{twoexits} shows that, in simple graphs, two exits suffice to capture the computational complexity of \textsc{Nemesis}.
Note however that this reduction does not, in general, preserve planarity.


\section{Polynomial Time Games} \label{sec:poly}

For some classes of instances, the winner of the game can be determined efficiently by using classical graph structures. We now investigate several such cases.

\subsection{Trees and Graphs of Maximum Degree~3}

Before addressing the simplest graph classes, we introduce a graph simplification procedure.

\paragraph*{Graph Simplification}

\begin{figure}[ht]
  \begin{center}
    \includegraphics[width=\textwidth]{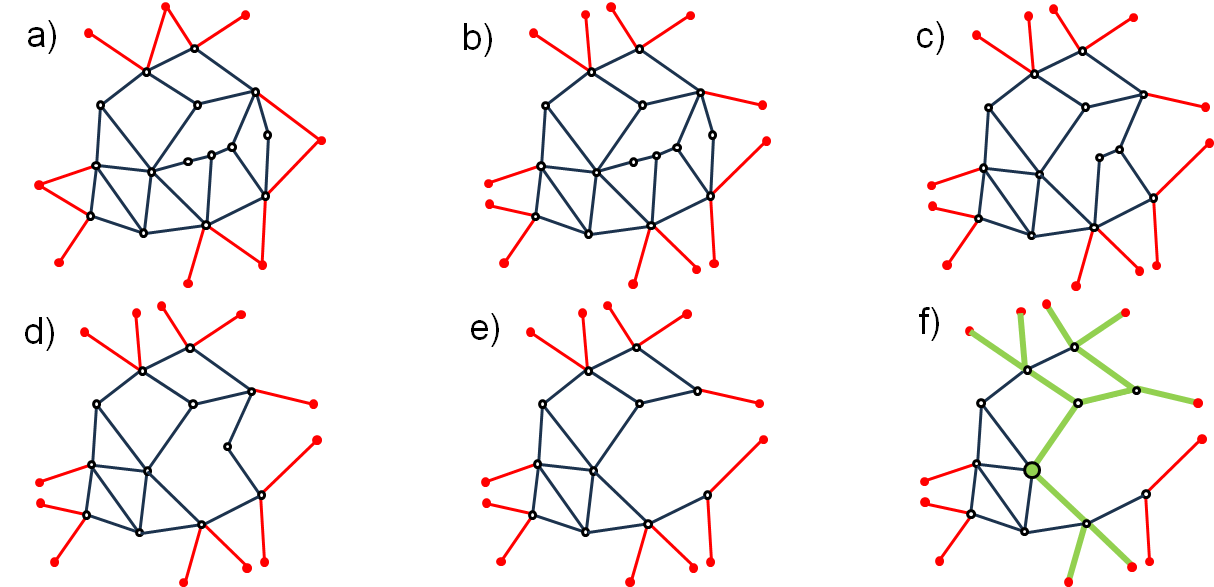}
  \end{center}
  \caption{\label{simpli} \textbf{Simplification of an input graph.}
  From~a) to~b), exits are duplicated.
  From~b) to~e), regular vertices of degree~$1$ or~$2$ are pruned.
  In~f), a binary escape tree.}
\end{figure}

Given a graph with exits and a starting vertex~$s$, we introduce a first reduction performed in two phases (Fig.~\ref{simpli}).  
The first phase consists in duplicating the exits so that each exit has degree~$1$. See for instance the transition from~a) to~b) in Fig.~\ref{simpli}.
This transformation simplifies the structure by avoiding spurious cycles passing through exits, and it preserves the outcome of the game.

The second transformation consists in removing from the graph all vertices such that, if the fugitive ever reaches one of them, he is inevitably doomed to lose.
More precisely, we iteratively prune all regular vertices of degree~$1$ or~$2$ (except the starting vertex~$s$), since such vertices act as traps for the fugitive.
As these vertices never play any role in a winning strategy, this pruning phase does not change the outcome of the game.

This pruning can be implemented using a Breadth-first Search (BFS) initialized with the initial set of vertices to be removed.
We also discard any connected component that does not contain~$s$.
Overall, this preprocessing runs in linear time.

From now on, we therefore assume that all input graphs have been previously simplified.
Note that this simplification never increases the degree of any vertex.

\paragraph*{Binary Escape Trees}

A key structure guaranteeing a win for the fugitive is the notion of a \textit{binary escape tree}.

\begin{definition}
A \emph{binary escape tree} of a graph~$G$ (given with a set of exits) is a rooted full binary tree (i.e., a rooted tree in which every internal node has exactly two children) whose vertices and edges belong to~$G$, and whose leaves are exits.
\end{definition}


Binary escape trees provide the fugitive with a simple winning strategy.

\begin{lemma}\label{Lem}
Given an instance of \textsc{Nemesis} with graph~$G=(V,E)$, a set of exits, and a starting vertex~$s$, if $s$ or one of its neighbors in~$G$ is the root of a binary escape tree, then the fugitive has a winning strategy.
\end{lemma}

\begin{proof}
The winning strategy is straightforward.
If the root of the binary escape tree is a neighbor of~$s$, the fugitive moves to this root in the first round.
Otherwise, if the tree is rooted at~$s$, he moves to one of its children.

At each round, the Nemesis can delete at most one edge.
Consequently, from the fugitive’s current position, at least one of the two subtrees of the current root remains intact.
At the next round, the fugitive moves to the root of this unaltered subtree.
Repeating this strategy ensures that after each of his moves, the fugitive occupies the root of a binary escape tree in the current graph.
Since the height of the binary escape tree strictly decreases at each round, the fugitive eventually reaches an exit and wins the game.
\end{proof}

We call the \textit{tree condition} the property that at least one vertex at distance at most~$1$ from the fugitive’s current position is the root of a binary escape tree.
Lemma~\ref{Lem} can then be restated as follows: if the tree condition is satisfied, the fugitive wins.

It is natural to ask whether this condition is not only sufficient but also necessary.
If this were the case, it would yield a simple characterization of the starting positions from which the fugitive has a winning strategy.
However, this is not true in general.
Indeed, there exist instances in which the fugitive wins even though the tree condition is not satisfied.
A counter-example occures with square grids.

\paragraph*{Square Grids}

\begin{figure}[ht]
  \begin{center}
    \includegraphics[width=\textwidth]{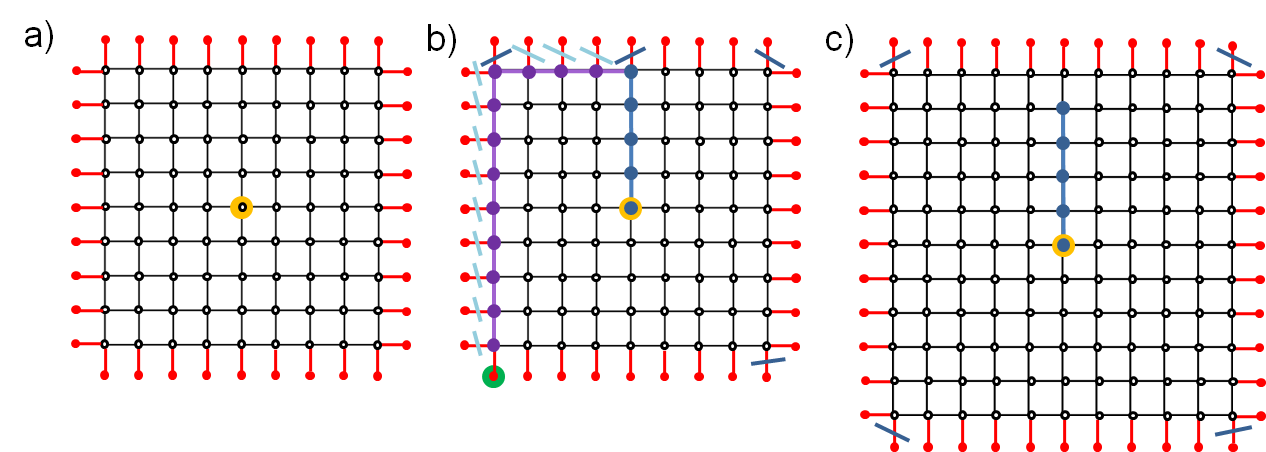}
  \end{center}
  \caption{\label{grids} \textbf{\textsc{Nemesis} on a square grid.}
  In~a), the input graph with the starting vertex in orange and exits in red.
  In~b), with a starting vertex at distance~$5$ from the closest exit, the fugitive wins by moving around the grid.
  In~c), with a starting vertex at distance~$6$ from the closest exit, the Nemesis wins by cutting one exit at each corner.}
\end{figure}

Square grids deserve special attention, as readers familiar with Conway’s Angel game may recognize in \textsc{Nemesis} a variant of the demon’s strategy \cite{Angel1,Angel2}.
The set of regular vertices of our instance forms an $m \times n$ square grid.
Each corner vertex is connected to two exits, while each other boundary vertex is connected to exactly one exit.
The instance and the corresponding strategies are illustrated in Fig.~\ref{grids}.

It is easy to see that if the fugitive starts at a vertex at distance at least~$6$ from the closest exit, then the Nemesis has a winning strategy by cutting the corner exits.
Conversely, if the distance from the starting vertex to the boundary vertices is at most~$5$, the fugitive has a winning strategy by reaching the boundary and then moving around the grid (we do not formally prove this claim).

An interesting feature of the grid is that when the starting vertex~$s$ is at distance~$5$ from the exits, the fugitive has a winning strategy, even though $s$ does not satisfy the tree condition.
We now prove this claim by contradiction.

Assume that a binary escape tree is rooted at a vertex at distance at most~$1$ from~$s$.
Then the distance from this root to the exits is at least~$4$.
This root has two children at distance at least~$3$ from the exits, each being the root of an independent binary escape tree.
Each of these two nodes in turn has two children at distance at least~$2$ from the exits, again serving as roots of independent binary escape trees.
At the next level, we therefore obtain $8$ nodes at distance at least~$1$ from the exits, each being the root of a binary escape tree.

However, any binary escape tree that is not rooted at an exit must contain at least one vertex adjacent to two exits.
Hence, this construction implies the existence of $8$ distinct vertices each adjacent to two exits.
Since the grid contains only $4$ such vertices, we obtain a contradiction.
Therefore, $s$ does not satisfy the tree condition.

\paragraph*{Tree Condition is Necessary and Sufficient for Trees and Graphs of Maximum Degree~3}
\label{sec-treeConditionLinear}

For arbitrary graphs, the tree condition is sufficient but not necessary to guarantee a win for the fugitive.
We now show that when the graph $G$ is either a tree or has maximum degree at most~$3$ (after graph simplification), this condition is also necessary.

\begin{lemma}\label{Lem2}
Given a \textsc{Nemesis} instance with a set of exits and a graph $G=(V,E)$ that is either a tree or has maximum degree at most~$3$ after graph simplification, the fugitive has a winning strategy from the starting position~$s$ if and only if $s$ or one of its neighbors in~$G$ is the root of a binary escape tree.
\end{lemma}


\begin{proof}
Note that the simplified graph of a tree or a graph of max degree $3$ is itself a tree or a graph of max degree $3$. The condition that the graph has been previously simplified does not reduce the scope of Lemma \ref{Lem2} but it is necessary to avoid artificial cycles which might pass through the exits in the initial graph before the duplication step. 

One direction of the equivalence is given by Lemma~\ref{Lem}. We prove the converse, namely that if the tree condition is not satisfied, then Nemesis has a winning strategy.

We proceed by induction on the size $n = \vert V \vert$ of the graph $G$.

We start with trees.
Our induction hypothesis is that for any tree with at most $n$ vertices, the fugitive wins if and only if one of the vertices at distance at most $1$ from his current position is the root of a binary escape tree.

Consider now a tree with $n+1$ vertices, and assume that the fugitive has a winning strategy from $s$. By Remark~\ref{noloop}, we may assume that this strategy never visits the same vertex twice. At his first move, the fugitive moves from $s$ to a vertex $v$ and never returns to $s$. Thus, the fugitive has a winning strategy in the subtree of $G$ rooted at $v$, whose size is at most $n$.

By the induction hypothesis, either $v$ itself or one of its neighbors is the root of a binary escape tree in this subtree. If $v$ is such a root, then the tree condition holds at $s$. Otherwise, at least one child of $v$ has this property.

Suppose that exactly one child $v'$ of $v$ is the root of a binary escape tree in this subtree. Then, immediately after the fugitive moves from $s$ to $v$, Nemesis can delete the edge $(v,v')$, thereby destroying the tree condition at $v$. By the induction hypothesis, the fugitive then has no winning strategy, contradicting our assumption. Hence, $v$ must have at least two children that are roots of binary escape trees.

Since $G$ is a tree, these two subtrees are vertex-disjoint. It follows that $v$ itself is the root of a binary escape tree. Thus, the tree condition is satisfied at $s$ through its neighbor $v$.

We now consider graphs of maximum degree $3$.
The proof follows the same inductive framework, but requires an additional final argument.

The induction hypothesis is that for any graph of maximum degree at most $3$ with at most $n$ vertices, the fugitive wins if and only if the tree condition is satisfied. Let $G$ be such a graph with $n+1$ vertices, and let $s$ be a winning starting position for the fugitive.

As before, we may assume that the fugitive follows a winning strategy without loops. Let $v$ be the first vertex reached by the fugitive from $s$ according to the winning strategy. After the first round (one move by the fugitive and one deletion by Nemesis), the remaining graph has size at most $n$, and the fugitive is at $v$. By the induction hypothesis, $v$ or one of its neighbors is the root of a binary escape tree.

If $v$ itself is such a root, the tree condition holds and we are done. Otherwise, let $v'$ be a neighbor of $v$ that is the root of a binary escape tree. Since, in this case, $v$ is regular, $v$ has two other neighbors: $s$, $v'$ and  $v''$ (after simplification the unique vertex of degree $2$ might be $s$).

Nemesis may delete the edge $(v,v')$ at the first round, forcing the fugitive to move toward $v''$. If $v''$ is not the root of a binary escape tree, then by the induction hypothesis the fugitive loses, contradicting our assumption. Hence, both $v'$ and $v''$ must be roots of binary escape trees.

If the two binary escape trees rooted at $v'$ and $v''$ are vertex-disjoint, then $v$ is itself the root of a binary escape tree, and the tree condition holds. Otherwise, these two trees share vertices. We now show that such a configuration contradicts the initial assumption that the fugitive has a winning strategy starting by the move $(s,v)$.

Since the graph has maximum degree $3$ and the fugitive never revisits a vertex, whenever the fugitive arrives at a vertex, one incident edge has already been used and will not be crossed again. Of the remaining two edges, Nemesis can delete one, thereby forcing the fugitive’s next move. 
If the two binary escape trees rooted at $v'$ and $v''$ are not vertex-disjoint, then their union contains a cycle~$C$ passing through $v$, $v'$, and $v''$, in which Nemesis can trap the fugitive. Indeed, she only has to remove, at each round, the single edge escaping from cycle $C$. As the graph is assumed to be simplified, the cycle $C$ does not contain exits. It ensures the win of the Nemesis and contradicts the assumption that the fugitive can escape after initially moving to $v$. 
We conclude that whenever the fugitive has a winning strategy in a tree or in a graph of maximum degree~$3$, the tree condition must be satisfied.

\end{proof}

The tree condition can be checked in linear time for trees using a BFS starting from $s$.
For graphs of maximum degree~$3$, the condition can also be checked in linear time, although the argument is slightly more involved.
The algorithm also performs a BFS from the starting vertex~$s$ and labels each visited vertex according to the first neighbor of~$s$ through which it is reached.
Since $\deg(s)\leq 3$, there are at most three such branches.

If a vertex is reached from two different branches, then both branches are disabled, as they necessarily share a cycle and cannot both belong to a binary escape tree.
If a vertex is reached twice from the same branch, this branch is disabled as well.

After the traversal, the vertex~$s$ is the root of a binary escape tree if and only if at least two branches remain enabled.
The correctness of the algorithm follows from the fact that in graphs of maximum degree~$3$, any branch containing a cycle cannot be part of a binary escape tree.
Conversely, due to the preliminary simplification, all the regular vertices are of degree $3$. Then any branch which does not contain a cycle is a binary exit tree. Hence, the existence of two enabled branches i.e without cycles implies that $s$ is the root of a binary escape tree.

\subsection{Blizzard}

We now investigate the complexity of \textsc{Blizzard} and prove Theorem~\ref{t2}. 
The game \textsc{Blizzard} differs from \textsc{Nemesis} by a restriction on the adversary’s power. 
The Nemesis, now called the \emph{Storm}, can delete only edges that are adjacent to the current position of the fugitive, now called the \emph{trapper}.
This restriction is crucial, as it leads to a much simpler characterization of the trapper’s winning positions than in \textsc{Nemesis}.

\paragraph*{Sets of Winning Positions}

In this subsection, it is more convenient to place us at the half-time of a round, after the trapper's move and before the Storm removes of an edge. 

We begin by defining a base set $W_0$ of winning positions as the set of exit vertices. Clearly, if the trapper stands at an exit, it means that he already won.
We then define $W_1$ as the set of regular vertices that have at least two exit edges, that is, two edges leading to vertices in $W_0$. If, at the half-time of a round, the trapper is at such a vertex and these edges have not been deleted, then the Storm can delete at most one of them, leaving a remaining path to an exit. Hence the trapper still wins.

We extend this construction inductively by defining the sets of vertices $W_k$, which we call the \emph{winning positions of rank $k$}. Although the terminology suggests that these positions correspond to winning strategies for the trapper, for the moment we define them purely inductively as follows:
for $k \geq 1$, the set $W_k$ consists of the vertices that have at least two neighbors in $\bigcup_{i<k} W_i$. The induction process continues until no new vertices are added, that is, until some $W_\ell=\emptyset$. We define the set of winning positions
$
W = \bigcup_{k < \ell} W_k .$

The sets $W$ can be computed by a BFS by progressively labeling the vertices at distance $1$ from $W$, and therefore in linear time with respect to the size of the graph.
We now justify the terminology by showing that $W$ indeed coincides with the set of winning positions for the trapper.

\begin{theorem}\label{t5}
In \textsc{Blizzard}, the trapper has a winning strategy if and only if his starting position is adjacent to a vertex of $W$.
\end{theorem}

\begin{proof}
For this proof, we still consider the position of the trapper at the half-time of a round, that is, after his move and before the Storm deletes an edge.

We first prove that if, at the half-time of a round, the trapper stands at a vertex belonging to $W$, then he has a winning strategy.
We proceed by induction on the rank $k$. We assume that for some $d \geq 0$, for every vertex in $\bigcup_{k \leq d} W_k$, the trapper has a winning strategy starting from this vertex, using only edges whose endpoints belong to $\bigcup_{k \leq d} W_k$.
Now, let $u \in W_{d+1}$.
By definition of $W_{d+1}$, the vertex $u$ has at least two neighbors $v$ and $v'$ in $\bigcup_{k \leq d} W_k$.
By the induction hypothesis, both $v$ and $v'$ are winning positions for the trapper, with winning strategies entirely contained in $\bigcup_{k \leq d} W_k$.
In \textsc{Blizzard}, the Storm can delete at most one edge per round, and this edge must be incident with the current position of the trapper.
Consequently, the Storm may delete either $(u,v)$ or $(u,v')$, but not both.
Moreover, since the winning strategies from $v$ and $v'$ do not rely on edges incident with $u$, deleting an edge incident with $u$ does not affect these strategies.
Therefore, starting from $u$, the trapper moves to either $v$ or $v'$.
If the Storm deletes the edge to one of them, the trapper moves to the other.
In both cases, he reaches a vertex from which he can force a win by the induction hypothesis.
This shows that, when standing at $u$ at the half-time of a round, the trapper has a winning strategy, thus extending the induction hypothesis to $d+1$.

We now prove the converse.
Assume that, at the half-time of a round, the trapper stands at a vertex $u \notin W$.
By construction of the sets $W_k$, such a vertex has at most one neighbor in $W$.
The Storm can therefore delete the unique edge (if any) connecting $u$ to $W$, and hence prevent the trapper from entering $W$.
Since $W_0 \subseteq W$ and $W_0$ is the set of exits, the trapper can never reach an exit.
Thus, in this case, the Storm has a winning strategy.
\end{proof}

Theorem \ref{t2} stating that \textsc{Blizzard} can be solved in linear time is a corollary of Theorem~\ref{t5} and the fact that the set of the Winning positions can be computed by a BFS.

\section{Hardness Results} \label{sec:hard}

In this section, we prove the hardness results of the paper.
We begin by proving Theorem~\ref{antoine} which establishes the complexity of deciding whether a given vertex is the root of a binary escape tree (Section~\ref{proofantoine}).
We then present a reduction from \textsc{Planar SAT} to \textsc{Nemesis} on planar multigraphs, proving Theorem~\ref{t3} (Section~\ref{red1}).
Finally, we prove with Theorem~\ref{t4} the PSPACE-completeness of \textsc{Nemesis} on arbitrary graphs (Section~\ref{pspace2}) using arbitrary multigraphs as an intermediary lemma (Section~\ref{pspace1}). As a corollary, we also derive the PSPACE-completeness of the \textsc{Cat Herding} problem (Section~\ref{pspace3}).

\subsection{NP-completeness of the Search for a Binary Escape Tree}\label{proofantoine}

\Cref{antoine}, restated below, shows the complexity gap of the binary escape tree problem, which was shown in Section~\ref{sec-treeConditionLinear} to be solvable in linear time in trees and graphs of maximum degree~3.

\binaryexit*

\begin{proof}
	The problem is clearly in NP: given a certificate (a subtree rooted at $r$), we can verify in polynomial time if it is a binary escape tree.
	To prove completeness, we reduce from \ltsat, which is defined as follows:
	
	\medskip
	\noindent
	\textbf{\ltsat}
	\newline
	\textbf{Input:} A Boolean formula $\varphi$ in conjunctive normal form such that each clause intersects at most one other clause, in which case they share at most one literal (called an \lsat formula).
	\newline
	\textbf{Question:} Is $\varphi$ satisfiable?
	\medskip
	
	\ltsat is known to be NP-complete~\cite{arkin2018selecting}.
	Note that, in an \lsat formula, each variable appears at most four times: at most twice as a positive literal and at most twice as a negative literal.
	
	Let $\varphi$ be an \lsat formula with variables $x_1, \ldots, x_n$ and clauses $c_1, \ldots, c_m$.
	We construct a graph~$G$ from $\varphi$ as follows.
	
	\medskip
	\noindent\textbf{Variable gadget.}
	For each variable $x_i$, we construct the following rooted tree (depicted on Fig.~\ref{fig-variableGadget}). The root $X_i$ has one exit child and two children $x_i$ and $\no{x_i}$. The node $x_i$ (resp. $\no{x_i}$) has two children $x_i^1$ and $x_i^2$ (resp. $\no{x_i}^1$ and $\no{x_i}^2$), each having two exit children. The idea behind the gadget is to allow the selection of one branch: selecting the $x_i$ (resp. $\no{x_i}$) branch corresponds to setting the variable $x_i$ as true (resp. false).
	
	\medskip
	\noindent\textbf{Clause gadget.}
	For each clause $c_j$, we construct the following simple gadget: two adjacent vertices $C_j^1$ and $C_j^2$, each having an exit child. 
	This gadget is depicted in Fig.~\ref{fig-clauseGadget}.
	
	\medskip
	\noindent\textbf{Connecting the gadgets.}
	For each $1\le j\le m$, we associate $C_j^1$ with the first literal of the clause $c_j$, and $C_j^2$ with its second and third literals. We connect $C_j^1$ and $C_j^2$ to the \textbf{opposite} associated literals of $c_j$ in the following way: let $x_i$ be the first literal in $c_j$, if $x_i$ already appeared in a previous\footnote{in the sense of the total order $c_1,\ldots,c_m$} clause (which is the only clause that $c_j$ intersects), we connect $C_j^1$ to $\no{x_i}^2$, otherwise we connect $C_j^1$ to $\no{x_i}^1$. The same is done for the second and third literals with $C_j^2$. For example, let $c_6 = x_1 \lor \no{x_3} \lor x_8$ where $x_8$ is the only literal of $c_6$ which appeared in a previous clause, we connect $C_6^1$ to vertex $\no{x_1}^1$ and $C_6^2$ to vertices $x_3^1$ and $\no{x_8}^2$, as depicted on Fig.~\ref{fig-connectingGadgets}.
	
	\begin{figure}[h]
		\centering
		\begin{subfigure}{0.38\linewidth}
			\centering
			\scalebox{0.85}{
				\begin{tikzpicture}
					\node[noeud] (X) at (2.75,3) {$X_i$};
					\node[noeud] (x) at (1.25,2) {$x_i$};
					\node[noeud] (x1) at (0.5,1) {$x_i^1$};
					\node[noeud] (x2) at (2,1) {$x_i^2$};
					\node[noeud] (nx) at (4.25,2) {$\no{x_i}$};
					\node[noeud] (nx1) at (3.5,1) {$\no{x_i}^1$};
					\node[noeud] (nx2) at (5,1) {$\no{x_i}^2$};
					\node[exit] (exit1) at (0,0) {};
					\node[exit] (exit2) at (1,0) {};
					\node[exit] (exit3) at (1.5,0) {};
					\node[exit] (exit4) at (2.5,0) {};
					\node[exit] (exit5) at (3,0) {};
					\node[exit] (exit6) at (4,0) {};
					\node[exit] (exit7) at (4.5,0) {};
					\node[exit] (exit8) at (5.5,0) {};
					\node[exit] (exit0) at (3.75,3) {};
					\draw (nx)to(X)to(x);
					\draw (x1)to(x)to(x2);
					\draw (nx1)to(nx)to(nx2);
					\draw (exit1)to(x1)to(exit2);
					\draw (exit3)to(x2)to(exit4);
					\draw (exit5)to(nx1)to(exit6);
					\draw (exit7)to(nx2)to(exit8);
					\draw (X)to(exit0);
				\end{tikzpicture}
			}
			\caption{The variable gadget for $x_i$.}
			\label{fig-variableGadget}
		\end{subfigure}
		\hfil
		\begin{subfigure}{0.15\linewidth}
			\centering
			\scalebox{0.85}{
				\begin{tikzpicture}
					\node[noeud] (C1) at (0,1) {$C_j^1$};
					\node[noeud] (C2) at (0,0) {$C_j^2$};
					\node[exit] (exit1) at (1,1) {};
					\node[exit] (exit2) at (1,0) {};
					\draw (exit1)to(C1)to(C2)to(exit2);
				\end{tikzpicture}
			}
			\caption{The clause gadget for $c_j$.}
			\label{fig-clauseGadget}
		\end{subfigure}
		\hfil
		\begin{subfigure}{0.42\linewidth}
			\centering
			\scalebox{0.85}{
				\begin{tikzpicture}
					\node[noeud] (nx1) at (0.5,1) {$\no{x_1}^1$};
					\node[noeud] (x3) at (2,1) {$x_3^1$};
					\node[noeud] (nx8) at (3.5,1) {$\no{x_8}^2$};
					\node[noeud] (C61) at (5,1.5) {$C_6^1$};
					\node[noeud] (C62) at (5,0.5) {$C_6^2$};
					\node[exit] (exit1) at (0,0) {};
					\node[exit] (exit2) at (1,0) {};
					\node[exit] (exit3) at (1.5,0) {};
					\node[exit] (exit4) at (2.5,0) {};
					\node[exit] (exit5) at (3,0) {};
					\node[exit] (exit6) at (4,0) {};
					\node[exit] (exit7) at (6,1.5) {};
					\node[exit] (exit8) at (6,0.5) {};
					\draw (exit1)to(nx1)to(exit2);
					\draw (exit3)to(x3)to(exit4);
					\draw (exit5)to(nx8)to(exit6);
					\draw (exit7)to(C61)to(C62)to(exit8);
					\draw[dashed] (nx1)to(0,2);
					\draw[dashed] (x3)to(1.5,2);
					\draw[dashed] (nx8)to(3,2);
					\draw[bend right,line width=0.3mm] (C61)to(nx1);
					\draw[bend left,line width=0.3mm] (C62)to(x3);
					\draw[bend left,line width=0.3mm] (C62)to(nx8);
				\end{tikzpicture}
			}
			\caption{Connecting the gadgets (connections are shown with bold edges). The clause is $c_6 = x_1 \lor \no{x_3} \lor x_8$, $x_8$ has already appeared in another clause as a literal, contrary to $x_1$ and $\no{x_3}$.}
			\label{fig-connectingGadgets}
		\end{subfigure}
		\caption{\textbf{The gadgets.} Variable and clause gadgets of the reduction from \ltsat. Exit vertices are filled in red.}
		\label{fig-gadgets}
	\end{figure}
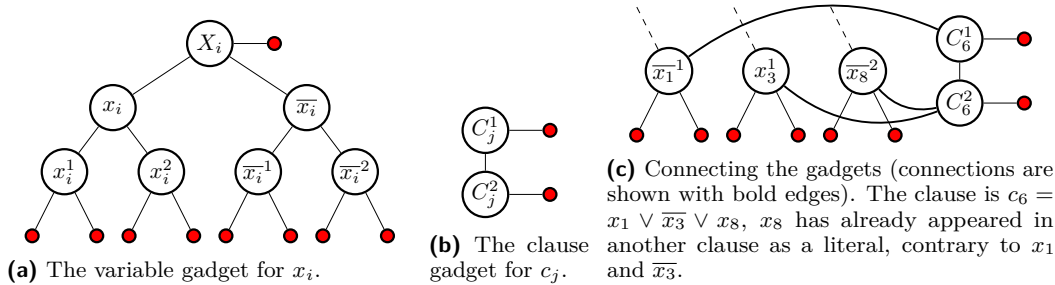
	
	\medskip
	\noindent\textbf{The complete construction.}
	We proceed as follows from formula $\varphi$. First, we have the root vertex $r$ forming a linear path with vertices $X'_1, \ldots, X'_n$, each linked to the next one and having a pendant variable gadget rooted at $X_1, \ldots, X_n$. Following $X'_n$, we then have a linear path of vertices $C'_1, \ldots, C'_m$, each linked to the next one and having a pendant clause gadget with $C_j^1$ connected to $C'_j$. Clause and variable gadgets are connected as described above. Finally, $r$ and $C'_m$ have adjacent exits\footnote{Note that the exit adjacent to $r$ can be removed to obtain a reduction to the problem of deciding whether there exists a binary escape tree with a root adjacent to the starting position.}. The full construction is depicted on Fig.~\ref{fig-reductionFromLinearSat}.
	
	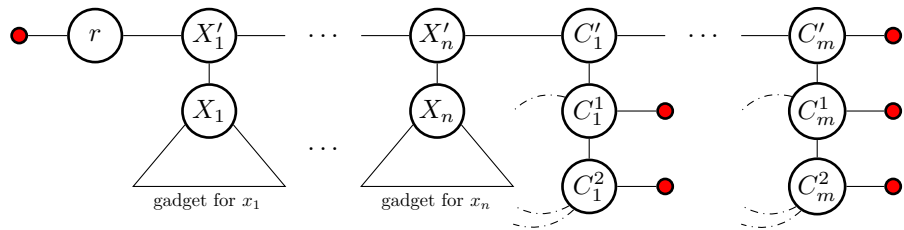
\begin{figure}[h]
		\centering
		\scalebox{1}{
			\begin{tikzpicture}
				\node[noeud] (r) at (0,1) {$r$};
				\node[exit] (exit0) at (-1,1) {};
				
				\node[noeud] (X'1) at (1.5,1) {$X'_1$};
				\node[noeud] (X1) at (1.5,0) {$X_1$};
				\node[text width=2.1cm,scale = 0.7] (var1) at (1.5,-1.2) {gadget for $x_1$};
				\draw (X1.210)to(0.5,-1)to(2.5,-1)to(X1.330);
				
				\draw (3,1) node {$\ldots$};
				\draw (3,-0.5) node {$\ldots$};
				
				\node[noeud] (X'n) at (4.5,1) {$X'_n$};
				\node[noeud] (Xn) at (4.5,0) {$X_n$};
				\node[text width=2.1cm,scale = 0.7] (var2) at (4.5,-1.2) {gadget for $x_n$};
				\draw (Xn.210)to(3.5,-1)to(5.5,-1)to(Xn.330);
				
				\node[noeud] (c1) at (6.5,1) {$C'_1$};
				\node[noeud] (c1a) at (6.5,0) {$C_1^1$};
				\node[noeud] (c1b) at (6.5,-1) {$C_1^2$};
				\node[exit] (ex1a) at (7.5,0) {};
				\node[exit] (ex1b) at (7.5,-1) {};
				
				\draw (8,1) node {$\ldots$};
				
				\node[noeud] (cm) at (9.5,1) {$C'_m$};
				\node[noeud] (cma) at (9.5,0) {$C_m^1$};
				\node[noeud] (cmb) at (9.5,-1) {$C_m^2$};
				\node[exit] (exma) at (10.5,0) {};
				\node[exit] (exmb) at (10.5,-1) {};
				
				\node[exit] (exit1) at (10.5,1) {};
				
				\draw (exit0)to(r)to(X'1);
				\draw (X'1)to(2.5,1);
				\draw (3.5,1)to(X'n);
				\draw (X'n)to(c1);
				\draw (c1)to(7.5,1);
				\draw (8.5,1)to(cm);
				\draw (cm)to(exit1);
				\draw (X'1)to(X1);
				\draw (X'n)to(Xn);
				\draw (c1)to(c1a);
				\draw (ex1a)to(c1a)to(c1b)to(ex1b);
				\draw (cm)to(cma);
				\draw (exma)to(cma)to(cmb)to(exmb);
				\draw[dash dot,bend right] (c1a)to(5.5,0);
				\draw[dash dot,bend left] (c1b)to(5.5,-1.5);
				\draw[dash dot,bend left] (c1b)to(5.5,-1.25);
				\draw[dash dot,bend right] (cma)to(8.5,0);
				\draw[dash dot,bend left] (cmb)to(8.5,-1.5);
				\draw[dash dot,bend left] (cmb)to(8.5,-1.25);
			\end{tikzpicture}
		}
		\caption{\textbf{The constructed graph.} The reduction from \ltsat. The variable gadgets are as in Fig.~\ref{fig-variableGadget}, and the dash-dot lines from the clause gadgets are connected to vertices in the variable gadgets as described in Fig.~\ref{fig-connectingGadgets}.}
		\label{fig-reductionFromLinearSat}
	\end{figure}
	
	The constructed graph $G$ has order $16n + 4m + 3$, and thus can be computed in polynomial time. Furthermore, $G$ has maximum degree~4. 
	We now prove that $\varphi$ is satisfiable if and only if $G$ has a binary escape tree rooted at $r$.
	
	\medskip
	\noindent ($\Rightarrow$) Pick a variable assignment that satisfies $\varphi$. From it, we derive a binary escape tree, depicted on Fig.~\ref{fig-treefromreduction}. The tree contains the path induced by the exit of $r$, $r,X'_1,\ldots,X'_n,C'_1,\ldots,C'_m$ and the exit of $C'_m$. To make it binary, we will fork in the gadgets. At each $X'_i$, we add to the tree, as a child of $X'_i$, the vertex $X_i$ and its exit child. If $x_i$ is set as true (resp. false), we also add the subtree of the variable gadget rooted at vertex $x_i$ (resp. $\no{x_i}$) as a child of $X_i$. At each $C'_j$, we add to the tree, as a child of $C'_j$, the vertex $C^1_j$ and its exit child. If the first literal of $c_j$ is set as true, we also add the subtree rooted in the variable gadget neighbor of $C_j^1$. Otherwise, we select one positive literal of $c_j$ (second or third one), and we add to the tree, as a child of $C_j^1$, the vertex $C_j^2$ along with its exit child and the subtree rooted in the variable gadget neighbor of $C_j^2$ associated with a positive literal of the clause $c_j$.
	This yields a binary escape tree: at each internal node, we are in one of three cases (1) two exit children (2) one exit child + one child root of a binary exit tree (3) two children roots of binary escape trees.
	
	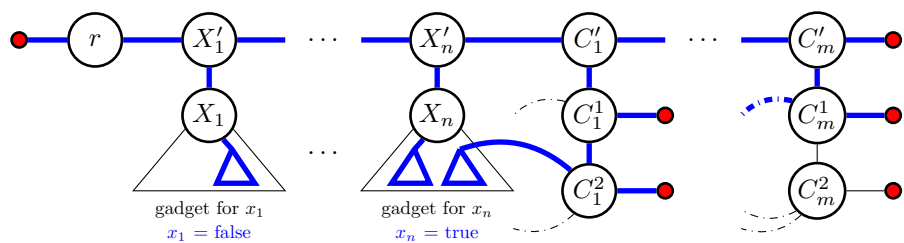
\begin{figure}[h]
		\centering
		\scalebox{1}{
			\begin{tikzpicture}
				\node[noeud] (r) at (0,1) {$r$};
				\node[exit] (exit0) at (-1,1) {};
				
				\node[noeud] (X'1) at (1.5,1) {$X'_1$};
				\node[noeud] (X1) at (1.5,0) {$X_1$};
				\node[text width=2.1cm,text centered,scale = 0.7] (var1) at (1.5,-1.4) {gadget for $x_1$\\\textcolor{blue}{$x_1 =$ false}};
				\draw (X1.210)to(0.5,-1)to(2.5,-1)to(X1.330);
				\draw[color = blue, line width = 2pt] (X1.300)to (1.8,-0.45)to(2.1,-0.9)to(1.6,-0.9)to(1.8,-0.45);
				
				\draw (3,1) node {$\ldots$};
				\draw (3,-0.5) node {$\ldots$};
				
				\node[noeud] (X'n) at (4.5,1) {$X'_n$};
				\node[noeud] (Xn) at (4.5,0) {$X_n$};
				\node[text width=2.1cm,text centered,scale = 0.7] (var2) at (4.5,-1.4) {gadget for $x_n$\\\textcolor{blue}{$x_n =$ true}};
				\draw (Xn.210)to(3.5,-1)to(5.5,-1)to(Xn.330);
				\draw[color = blue, line width = 2pt] (Xn.240)to (4.2,-0.45)to(3.9,-0.9)to(4.4,-0.9)to(4.2,-0.45);
				
				\node[noeud] (c1) at (6.5,1) {$C'_1$};
				\node[noeud] (c1a) at (6.5,0) {$C_1^1$};
				\node[noeud] (c1b) at (6.5,-1) {$C_1^2$};
				\node[exit] (ex1a) at (7.5,0) {};
				\node[exit] (ex1b) at (7.5,-1) {};
				
				\draw (8,1) node {$\ldots$};
				
				\node[noeud] (cm) at (9.5,1) {$C'_m$};
				\node[noeud] (cma) at (9.5,0) {$C_m^1$};
				\node[noeud] (cmb) at (9.5,-1) {$C_m^2$};
				\node[exit] (exma) at (10.5,0) {};
				\node[exit] (exmb) at (10.5,-1) {};
				
				\node[exit] (exit1) at (10.5,1) {};
				
				\draw[color = blue, line width = 2pt] (exit0)to(r)to(X'1);
				\draw[color = blue, line width = 2pt] (X'1)to(2.5,1);
				\draw[color = blue, line width = 2pt] (3.5,1)to(X'n);
				\draw[color = blue, line width = 2pt] (X'n)to(c1);
				\draw[color = blue, line width = 2pt] (c1)to(7.5,1);
				\draw[color = blue, line width = 2pt] (8.5,1)to(cm);
				\draw[color = blue, line width = 2pt] (cm)to(exit1);
				\draw[color = blue, line width = 2pt] (X'1)to(X1);
				\draw[color = blue, line width = 2pt] (X'n)to(Xn);
				\draw[color = blue, line width = 2pt] (c1)to(c1a);
				\draw[color = blue, line width = 2pt] (ex1a)to(c1a)to(c1b)to(ex1b);
				\draw[color = blue, line width = 2pt] (cm)to(cma);
				\draw[color = blue, line width = 2pt] (exma)to(cma);
				\draw (cma)to(cmb)to(exmb);
				\draw[dash dot,bend right] (c1a)to(5.5,0);
				\draw[dash dot,bend left] (c1b)to(5.5,-1.5);
				\draw[bend right,color = blue, line width = 2pt] (c1b)to(4.8,-0.45);
				\draw[color = blue, line width = 2pt] (4.8,-0.45)to(5.1,-0.9)to(4.6,-0.9)to(4.8,-0.45);
				\draw[dash dot,bend right,color = blue, line width = 2pt] (cma)to(8.5,0);
				\draw[dash dot,bend left] (cmb)to(8.5,-1.5);
				\draw[dash dot,bend left] (cmb)to(8.5,-1.25);
			\end{tikzpicture}
		}
		\caption{\textbf{Illustration of the equivalence.} Partial illustration of the binary escape tree obtained from a valid variable assignment. For example, variable $x_1$ is fixed as false, $x_n$ as true. The first literal of $c_1$ is false but the second one is true and equal to $x_n$. The first literal of $c_m$ is true.}
		\label{fig-treefromreduction}
	\end{figure}
	
	\medskip
	\noindent ($\Leftarrow$) Let $T$ be a binary exit tree of $G$ rooted at $r$. It satisfies the following properties (in the remainder of the proof, \emph{child} always refers to a child in $T$):
	\begin{enumerate}[label=(\roman*)]
		\item The induced path containing the exit of $r$, $r,X'_1,\ldots,X'_n,C'_1,\ldots,C'_m$ and the exit of $C'_m$ is in $T$.\\
		This comes from the fact that these vertices (except the root $r$) have exactly degree 3 and that $T$ is a full binary tree (except $r$, no vertex has degree 2 in $T$).
		\item The neighborhood of this induced path is also in $T$, namely all vertices $X_i$ and $C_j^1$. \\
		Indeed, each vertex of the path has exactly one neighbor not in the path, which forces $T$, as a binary tree, to contain all of those vertices.
		\item For each $1\le j\le m$, at least one of the two vertices $C_j^1$ and $C_j^2$ has a child belonging to a variable gadget. \\ Otherwise, one of those vertices would be of degree~1 in $T$, and the subtree rooted in $C_j^1$ would not be full binary, a contradiction.
		\item Each vertex $X_i$ has at least one of $x_i$ and $\no{x_i}$ as a child.\\
		Indeed, in $T$, each vertex $X_i$ must have two children. One child can be its neighbor exit. The second one is necessarily one of the vertices $x_i$ or $\no{x_i}$, together with their respective descendants.
		\item Vertices $x_i^1$ and $x_i^2$ (resp. $\no{x_i}^1$ and $\no{x_i}^2$) can be children of clause gadget vertices only if $\no{x_i}^1$ and $\no{x_i}^2$ (resp. $x_i^1$ and $x_i^2$) are not.\\
		Otherwise, if a clause gadget is adjacent in $T$ to both subtrees rooted in $x_i$ and $\no{x_i}$ respectively, then, from the argument used in (iv), $T$ contains a cycle, a contradiction.
	\end{enumerate}
	
	We now construct the following truth assignment for the variables: if, in $T$, $x_i$ is a child of $X_i$, set variable $x_i$ as true. Otherwise, set variable $x_i$ as false. We claim that properties (i)-(v) have the following direct translations:
	\begin{itemize}
		\item Each variable is assigned as true or false: (iv)\footnote{Note that some variables may have both $x_i$ and $\no{x_i}$ as children of $X_i$; this means that they are not used to satisfy clauses, and thus can be set as true or false indiscriminately (here, we chose true).}.
		\item Each clause is associated with at least one of its literals: (iii).
		\item Literals with which clauses are associated are set as true: otherwise $T$ contains a cycle.
		\item A variable cannot be associated to two clauses where it appears as opposite literals: (v). 
	\end{itemize}

	Those imply that the assignment is coherent and satisfies $\varphi$, completing the reduction.
\end{proof}



\subsection{Hardness of \textsc{Nemesis} for Planar Multigraphs}\label{red1}

We now prove Theorem \ref{t3}. Readers familiar with the reduction of \textsc{QSAT} to Geography~\cite{Geography} or to the Canadian Traveler Problem \cite{Papa} may recognize the general structure of the reduction from \textsc{QSAT} to graph problems. These classical reductions use a main central path around which  local choices encode the Boolean variable assignment of a \textsc{SAT} instance. 

\paragraph*{Planar Monotone Rectilinear 3SAT}
We do not directly reduce \textsc{QSAT} but rather the NP-complete problem \textsc{Planar Monotone Rectilinear 3SAT} \cite{PM3SAT}, which we denote by \textsc{PMR3SAT}. A key feature of these \textsc{3SAT} instances is that each clause is either all-positive (containing only non-negated literals) or all-negative (containing only negated literals). An important property of \textsc{Planar Monotone Rectilinear 3SAT} is that it admits a geometric representation by non-crossing line segments.

\begin{figure}[ht]
  \begin{center}{
		\includegraphics[width=0.55\textwidth]{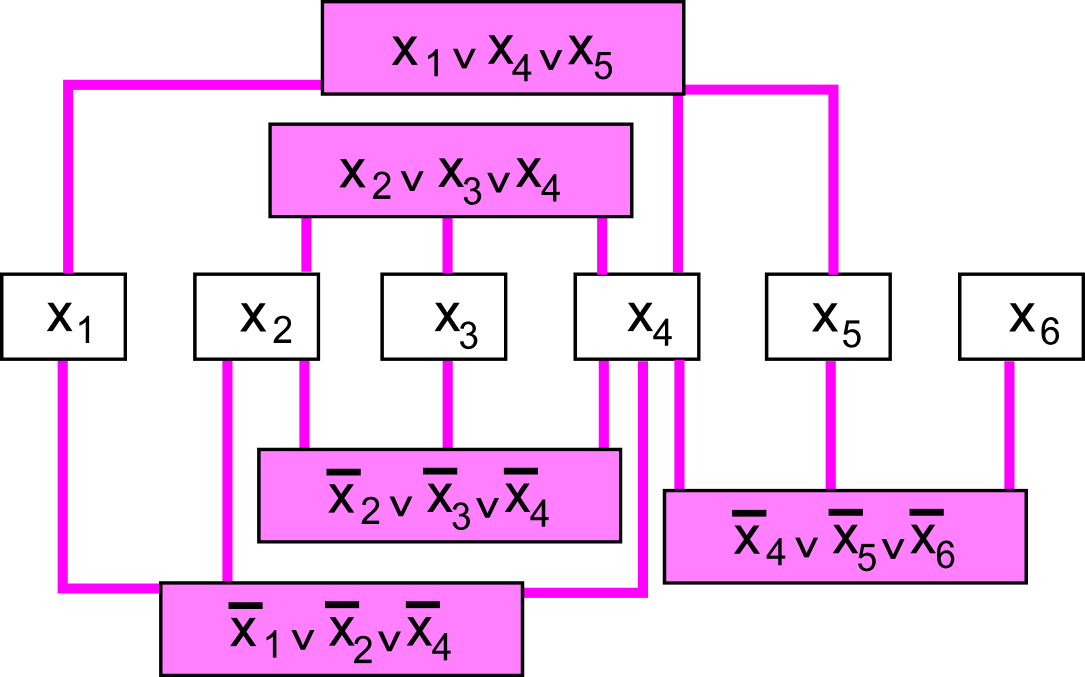}\vspace{0.5cm}
        \includegraphics[width=0.98\textwidth]{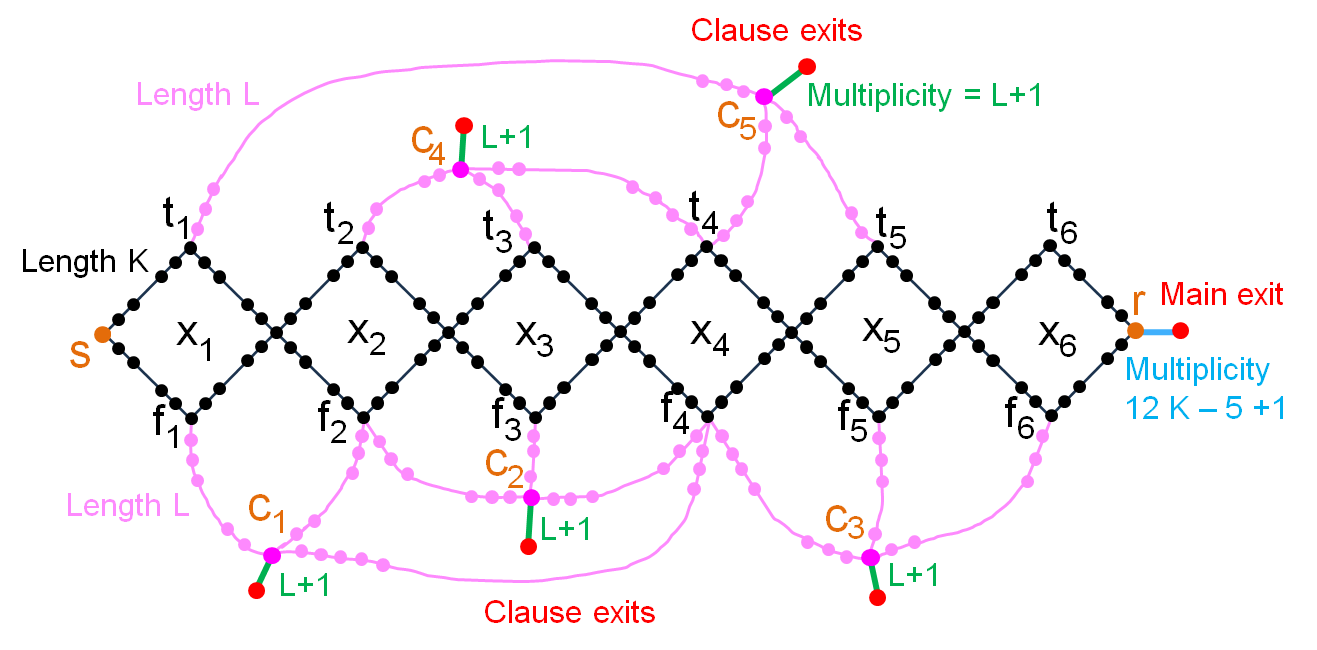}
  }
	\end{center}
	\caption{\label{reduction1} \textbf{Reduction of a \textsc{PMR3SAT} instance to a \textsc{Nemesis} instance.} We reduce the instance of \textsc{PMR3SAT} represented above. All the edges of the Nemesis instance are unbreakable except the edges adjacent to exits (in green). Their mulitplicity is $L+1$ for the clause exists and it is $12K-5+1$ for the main exit.}
\end{figure}

In this representation, Boolean variables appear as a sequence of axis parallel rectangles lying on the horizontal axis. Clauses correspond to axis parallel rectangles placed in the half-plane $y < 0$ for all-negative clauses, and in $y > 0$ for all-positive clauses. The participation of a variable in a clause is encoded by a vertical segment that connects the rectangle of the variable to that of the clause. An instance of \textsc{PMR3SAT} is shown in the upper part of Fig.~\ref{reduction1}. We do not rely on the fact that these segments are axis-parallel.

\paragraph*{Construction of the Nemesis instance}

We present the reduction of an instance of \textsc{PMR3SAT} to an instance of \textsc{Nemesis} in a planar multigraph. Let $n$ be the number of variables and $m$ the number of clauses of the \textsc{PMR3SAT} instance. The construction can be followed in Fig.~\ref{reduction1}.

Each Boolean variable is represented by a cycle of size $4K$, where $K$ is a constant chosen larger than the maximum number of clauses in which any given variable appears (larger than $m$ is sufficient). Thus, the ordered list of the $n$ variables of the \textsc{PMR3SAT} instance is encoded by a sequence of cycles - one per variable - each connected to the next by an intermediate vertex.

The key idea of the reduction is that, when traveling from the leftmost start vertex $s$ to the rightmost regular vertex $r$, the fugitive must traverse each cycle on one side or the other, so that his path from $s$ to $r$ encodes an assignment of the $n$ variables. For this purpose, we introduce intermediate vertices $t_i$ (for true) and $f_i$ (for false) on the two sides of each cycle. For instance, a path from $s$ to $r$ passing through vertices $t_1$, $f_2$, $t_3$, $t_4$, $f_5$, $f_6$ corresponds to the assignment where $x_1$, $x_3$ and $x_4$ are true, and $x_2$, $x_5$ and $x_6$ are false.

Observe that, regardless of the choice of side for each variable, all $s$-$r$ paths passing through the variable cycles have length $2nK$ (in the figure, with $n=6$, this length is $12K$).

Since this central path is crucial to the reduction, we ensure that the Nemesis cannot destroy its structure: all its edges are given multiplicities that make them unbreakable. The same applies to the additional edges introduced below. Almost all edges are therefore unbreakable. The only edges whose multiplicities allow the Nemesis to delete them are those incident with exits. The main exit is connected to the rightmost regular vertex $r$, and we set the multiplicity of this edge to $2nK - m + 1$. The remaining exits correspond to the clauses.

Each clause is represented by a vertex $c_i$ connected to an exit by an edge of multiplicity $L+1$, where $L$ is chosen larger than $2nK$. Each clause vertex $c_h$ is also connected to the cycles of its variables.

For an all-positive clause $x_i \vee x_j \vee x_k$, the vertex $c_h$ is linked to $t_i$, $t_j$, and $t_k$ by unbreakable paths of length $L$.

For an all-negative clause $\overline{x_i} \vee \overline{x_j} \vee \overline{x_k}$, the vertex $c_h$ is linked to $f_i$, $f_j$, and $f_k$ by unbreakable paths of length $L$.

The choice of $L$, which is larger than the distance from $s$ to $r$, guarantees that these clause paths cannot create shortcuts between $s$ and $r$.

\paragraph*{How does it work?}

We first consider the direct paths from $s$ to $r$. For each variable $x_i$, the path goes through either $t_i$ or $f_i$. Assume, without loss of generality, that the fugitive arrives at $t_i$. From there, he may attempt to use one of the corresponding clause paths to reach an exit. When the fugitive is at $t_i$, his distance to $c_h$ is $L$, and the initial multiplicity of the exit edge from $c_h$ is $L+1$. Thus, if this exit edge has not been attacked by the Nemesis at least once beforehand, the fugitive can reach the clause exit, since the Nemesis no longer has enough time to remove that edge.

This forces the Nemesis to anticipate: before the fugitive reaches $t_i$, he must traverse a path of length $K$. During these $K$ rounds, the Nemesis must attack the exit edges of all clauses connected to $t_i$. Reducing the multiplicity of each such exit edge by one is sufficient to make the corresponding clause exit harmless. This reduction is unavoidable, because otherwise the fugitive could escape through that exit. Consequently, for each clause reachable from the fugitive's path, the Nemesis must spend one round reducing an exit edge’s multiplicity from $L+1$ to $L$.

\paragraph*{Equivalence between the \textsc{PMR3SAT} and the corresponding \textsc{Nemesis} instances}\label{multi}
We now show that when the \textsc{PMR3SAT} instance is satisfiable, the fugitive has a winning strategy by following the path corresponding to a satisfying assignment. Along this path, the vertices $t_i$ and $f_i$ visited by the fugitive are connected to all $m$ clauses. Hence, the Nemesis must spend $m$ rounds preventing escape through the clause exits. This leaves only $2nK - m$ rounds during the fugitive’s traversal from $s$ to $r$ (a path of length $2nK$) to destroy the main exit edge. Since its multiplicity is set to $2nK - m + 1$, that edge still has multiplicity $1$ at the beginning of round $2nK - m + 1$ (after the fugitive reaches $r$), allowing the fugitive to exit.\\

We now show that if the \textsc{PMR3SAT} instance is not satisfiable, then the fugitive cannot escape. In this case, the fugitive’s path from $s$ to $r$ visits at most $m-1$ clauses, so the Nemesis has enough free rounds to remove the main exit edge.

Could the fugitive take a better path? Using a clause path is not advantageous: if he travels from $t_i$ or $f_i$ to $c_h$, then at every round the Nemesis can reduce the multiplicity of that clause's exit edge. The Nemesis has exactly enough time to finish removing the exit edge before the fugitive can traverse it. The fugitive must then return to the central path, during which the Nemesis has ample time to destroy the main exit and win.

Another possibility would be to backtrack within the variable cycles, but this would require the fugitive to visit some vertex twice. This cannot help: if the fugitive has a winning strategy at all, then he has one without loops (Remark \ref{noloop}).

Thus, regardless of the path chosen by the fugitive, he cannot reach an exit.

Let us check now that the reduction takes only a polynomial time. The number of vertices involved in the central path is lower than $4nK$ with $K\leq m$. The clause paths require $3(L+1)m$ vertices with $L$ of the order of $2nK$. Then the total number of vertices is $O(m^2 n^2)$. The \textsc{Nemesis} instance encoding the initial \textsc{PMR3SAT}  intance is thus computed in polynomial time. 

This completes the NP-hardness proof, showing Theorem \ref{t3}.

\subsection{PSPACE-completeness of \textsc{Nemesis} for Arbitrary Multigraphs}\label{pspace1}

We now prove an intermediate result toward Theorem~\ref{t4}, namely the PSPACE-completeness of \textsc{Nemesis} for multigraphs. 

\paragraph*{The \textsc{QSAT} game between the Universal and Existential Players}

To establish the PSPACE-hardness of \textsc{Nemesis}, we need to move from the previous reduction from \textsc{3SAT} (or \textsc{SAT} in general) to a reduction from \textsc{Quantified SAT} (\textsc{QSAT}). This amounts to going from a formula $\varphi$ with only existential quantifiers,
$$\exists x_1 , \exists x_2  , ... , \exists x_i , ... , \exists x_n ,  \varphi\left((x_i)_{1\leq i \leq n}\right)$$ 
to a formula with alternating existential and universal quantifiers,
$$\exists x_1 , \forall x_2  , ... , \exists x_{2i-1}  , \forall x_{2i}, ... , \forall x_{2n} ,  \varphi\left((x_i)_{1\leq i \leq 2n}\right).$$
Variables quantified existentially are called \textit{existential}, while the others are called \textit{universal}.

The satisfiability of such a quantified formula can be viewed as a game between an \textit{existential player}, who assigns values to the existential variables with the objective of satisfying the formula, and a \textit{universal adversary}, who assigns values to the universal variables with the objective of preventing satisfaction. It is this game that we reduce to \textsc{Nemesis} in order to prove PSPACE-hardness.

\paragraph*{The Electric Gadget and its Analysis}

We now consider the initial reduction of \textsc{SAT} presented in Section \ref{red1} and explain how to update it in order to reduce the \textsc{QSAT} game. The main idea is to give the Nemesis control on the fugitive's path over every universal variable, thereby emulating the power of the universal player. To explain the construction, we use an electrical analogy.

\begin{figure}[ht]
  \begin{center}{
		\includegraphics[width=0.95\textwidth]{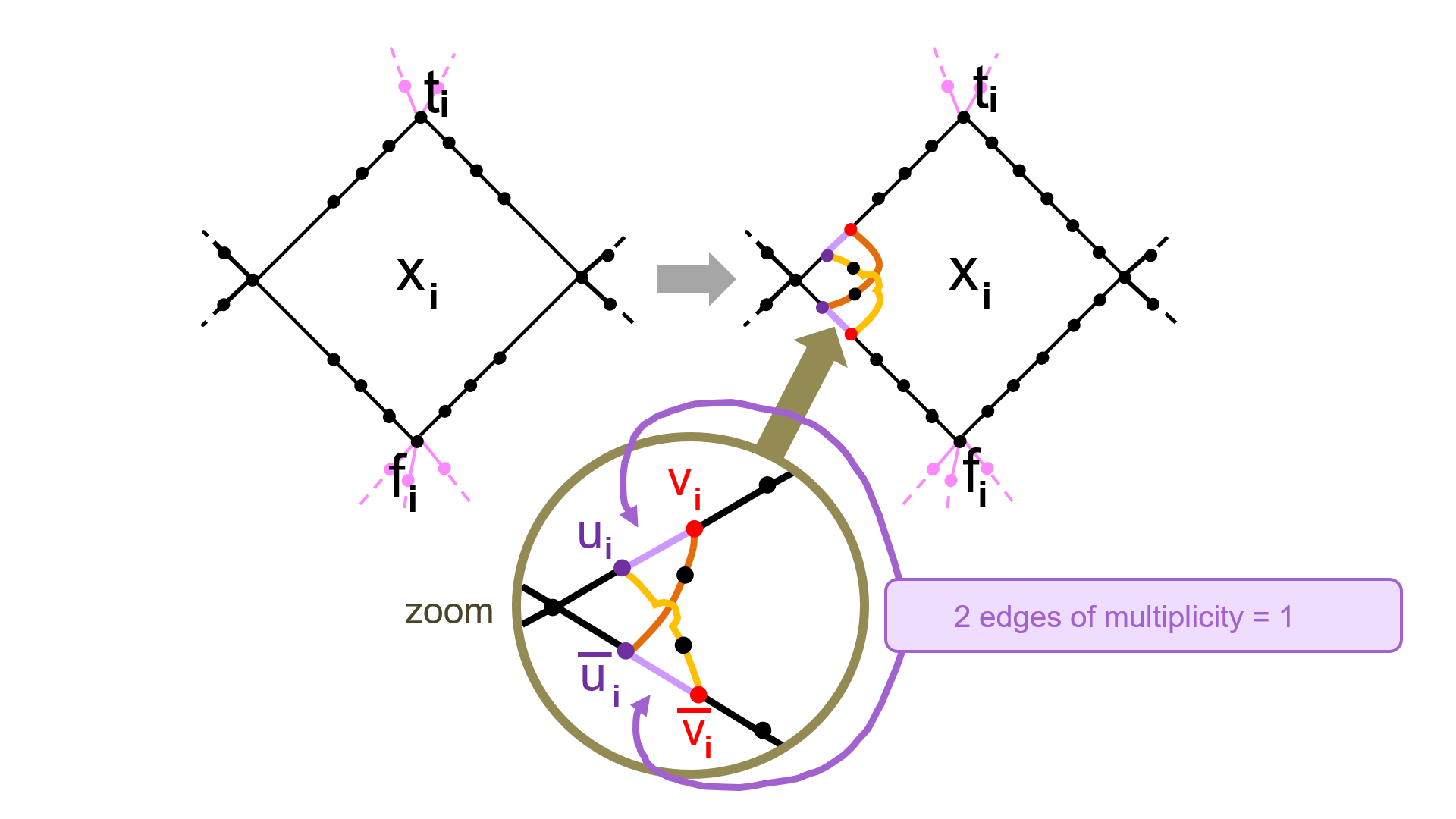}
  }
	\end{center}
	\caption{\label{gadget} \textbf{The Electric gadget to simulate a universal quantifier.} On the left, the initial variable gadget used to reduce \textsc{3SAT}. On the right, its update which allows the Nemesis to control the path followed by the fugitive.  }
\end{figure}

A first idea is to introduce, on each path corresponding to a universal variable (the $true$ path and the $false$ path), an edge of multiplicity $1$ that the Nemesis can use as a fuse. However, this naive approach potentially gives the Nemesis the ability to destroy both parallel fuses, which would allow her to win regardless of the fugitive’s choices. Thus, while this idea provides useful intuition, it grants too much power to the Nemesis and must be refined.
We therefore introduce a fuse on each path of the universal variables and complement it with two derivation paths. This construction yields the so-called \textit{electric gadget}, illustrated in Fig.~\ref{gadget}.

Let us now describe the gadget in its full details.
The gadget consists of a fuse (an edge of multiplicity one) placed on each branch immediately after the first edge of the paths corresponding to a universal variable $x_i$. The two fuses connect the vertices  $u_i$ and $v_i$ on the $true$ path, and vertices $\overline{u_i}$ and $\overline{v_i}$ on the $false$ path. We complete the electric gadget with two additional paths of length $2$: one from $\overline{u_i}$ to $v_i$, and the other from $u_i$ to $\overline{v_i}$. All edges on these two paths are unbreakable.

We now explain why this gadget gives the Nemesis control over the universal variable $x_i$. We first recall that, in the initial reduction from \textsc{SAT} to \textsc{Nemesis}, the race between the fugitive reaching the main exit and the Nemesis destroying the main exit edge is tight: if the fugitive loses even one single round, he is defeated.
We return to the electric gadget and assume, without loss of generality, that the Nemesis has chosen the value of variable $x_i$ to be true, meaning that she intends to forbid the fugitive from passing through $f_i$. When the fugitive reaches the cycle corresponding to the variable $x_i$, his first move must go either to $u_i$ or to $\overline{u_i}$. We analyze both cases.

\begin{itemize}
\item The fugitive moves to $u_i$. In this case, the Nemesis does not destroy the fuse $u_i v_i$. The fugitive then has two possible moves. If he goes to $v_i$, he effectively sets $x_i = true$, as expected by the Nemesis. If, on the contrary, he moves toward $\overline{v_i}$, he reaches $\overline{v_i}$ one round later than he would have by taking the direct path through $\overline{u_i}$. Since arriving at the main exit on time is a tight condition, this detour necessarily leads to defeat. Returning backwards would also result in defeat. Hence, the fugitive has no viable option other than proceeding toward $t_i$, as dictated by the Nemesis.
\item The fugitive moves to $\overline{u_i}$. Since the Nemesis has decided that the fugitive must go through $t_i$ rather than $f_i$, she immediately deletes the fuse between $\overline{u_i}$ and $\overline{v_i}$. The fugitive cannot turn back, as doing so would again condemn him. His only remaining option is to move to $v_i$, and thus toward $t_i$, as imposed by the Nemesis. This detour costs the fugitive one round, but the Nemesis has also spent one round deleting the fuse. These two rounds therefore cancel out, and, in terms of timing, the situation is equivalent to that in the original \textsc{SAT} reduction.
\end{itemize}

Another possible scenario is that the Nemesis deletes one or both fuses associated with variable $x_i$ before the fugitive reaches $u_i$ or $\overline{u_i}$. We now show that she cannot take advantage from such a strategy. By spending a round to delete at least one fuse prematurely, the Nemesis allows the fugitive to use either one of the unbreakable detours $u_i,\overline{v_i},f_i$ or $\overline{u_i},v_i,t_i$ without being condemned to lose. This gives the fugitive an advantage, making such a strategy suboptimal for the Nemesis.

This analysis shows that the electric gadget allows the Nemesis to choose whether the fugitive passes through $t_i$ or $f_i$, a choice that directly corresponds to the value of the universal variable $x_i$. This control comes at no net cost to the Nemesis. Of course, she may choose not to exercise this power, but refraining from doing so does not give her additional rounds to destroy the main exit edge. Any such restraint is therefore useless. The control granted by the electric gadget over the universal variable cannot be exploited elsewhere in the game.

\paragraph*{Proof of PSPACE hardness of \textsc{Nemesis} for Multigraphs} 

We now prove the PSPACE-hardness of \textsc{Nemesis} for multigraphs. The construction of the graph of \textsc{Nemesis} encoding a \textsc{QSAT} instance proceeds in two steps. First, we build the graph corresponding to the reduction from the \textsc{SAT} formula, as presented in Section~\ref{red1}. Second, we add the electric gadget at the entries of all universal Boolean variables.

This modification increases the size of the variable cycles by increasing the value of $K$ by a constant factor, but it does not otherwise affect the behavior of the construction.

We explore both directions of the reduction.

($\Rightarrow$) Assume that the quantified formula is satisfiable. This means that, in the corresponding \textsc{QSAT} game, the existential player has a winning strategy that satisfies all $m$ clauses. If the fugitive follows the same choices along the variable gadgets, he enforces the Nemesis to spend one round for each of the $m$ clauses of the \textsc{SAT} formula in order to block the clause exits.

For universal variables, the Nemesis may force the fugitive to lose one round by means of the electric gadget; however, in each such situation, the Nemesis must also spend one round to delete a fuse. Consequently, these rounds compensate each other, and the Nemesis gains no advantage. It follows that the fugitive reaches vertex $r$ before the Nemesis has completed the destruction of the main exit edge. At that point, this edge still has multiplicity $1$, and the fugitive escapes and wins the game. In other words, the fugitive has a winning strategy. 

($\Leftarrow$) Assume that the quantified formula is not satisfiable. Then, in the \textsc{QSAT} game, the universal player has a strategy that prevents the existential player from satisfying all clauses. By following an identical strategy, the Nemesis prevents the fugitive from being able to approach all $m$ clause exits: at least one clause is necessarily missed.
This missing clause yields one additional round that the Nemesis can devote to destroying the main exit edge. As a consequence, when the fugitive completes his path from $s$ to $r$, the edge to the main exit is no longer passable. Therefore, the Nemesis necessarily wins the game. It means that the Nemesis has a winning strategy.

This completes the proof of PSPACE-hardness of \textsc{Nemesis} for multigraphs.
To establish PSPACE-completeness, it remains to show that \textsc{Nemesis} belongs to PSPACE. Since \textsc{Nemesis} is a finite two-player game with perfect information, each game position can be encoded using polynomial space. Moreover, by Remark~\ref{noloop}, if the fugitive has a winning strategy, then he has one that does not visit the same vertex twice. Therefore, it is sufficient to consider plays without repeated vertices, and the length of any play is bounded by the number of vertices of the graph.
The winner of the game can thus be determined by a depth-first exploration of the game tree, evaluating positions recursively according to the current player. Although this procedure may require exponential time, it uses only polynomial space, which shows that \textsc{Nemesis} on multigraphs belongs to PSPACE.

Combining this result with the PSPACE-hardness, we conclude that \textsc{Nemesis} on multigraphs is PSPACE-complete.

\subsection{PSPACE-completeness of \textsc{Nemesis} for Arbitrary Graphs}\label{pspace2}

We now prove the PSPACE-completeness of \textsc{Nemesis} for arbitrary graphs (Theorem~\ref{t4}).
Rather than redoing the proof from scratch, we explain how the PSPACE-completeness proof for \textsc{Nemesis} on multigraphs can be adapted to obtain the same result for simple graphs.

Our objective is therefore to start from a \textsc{Nemesis} instance obtained by reducing an instance of \textsc{QSAT} and to construct an equivalent \textsc{Nemesis} instance whose underlying structure is a graph (\emph{i.e.}, without multiedges).

As a first step, we recall the key features of the multigraph instance produced by the \textsc{QSAT} reduction:
\begin{enumerate}
    \item The construction uses only four distinct values of edge multiplicities.
    Edges of multiplicity~$1$ appear in the electric gadgets.
    Exit edges have multiplicities $L+1$ and $2nK - m + 1$.
    All remaining edges are unbreakable: their multiplicity can be fixed to a value $N$ larger than the total number of rounds of the game, so that the Nemesis has no reason to attempt deleting them.
    This value $N$ is polynomial in the number of variables and clauses of the original \textsc{QSAT} instance.
    
    \item A crucial property of the reduction is that the fugitive’s winning path is tight.
    If the fugitive loses even a single round along his path from $s$ to $r$, then he can no longer reach $r$ before the Nemesis finishes destroying the main exit edge.
    This tightness is essential for the correctness of the reduction.
\end{enumerate}

\begin{figure}[ht!]
  \begin{center}{
        \includegraphics[width=\textwidth]{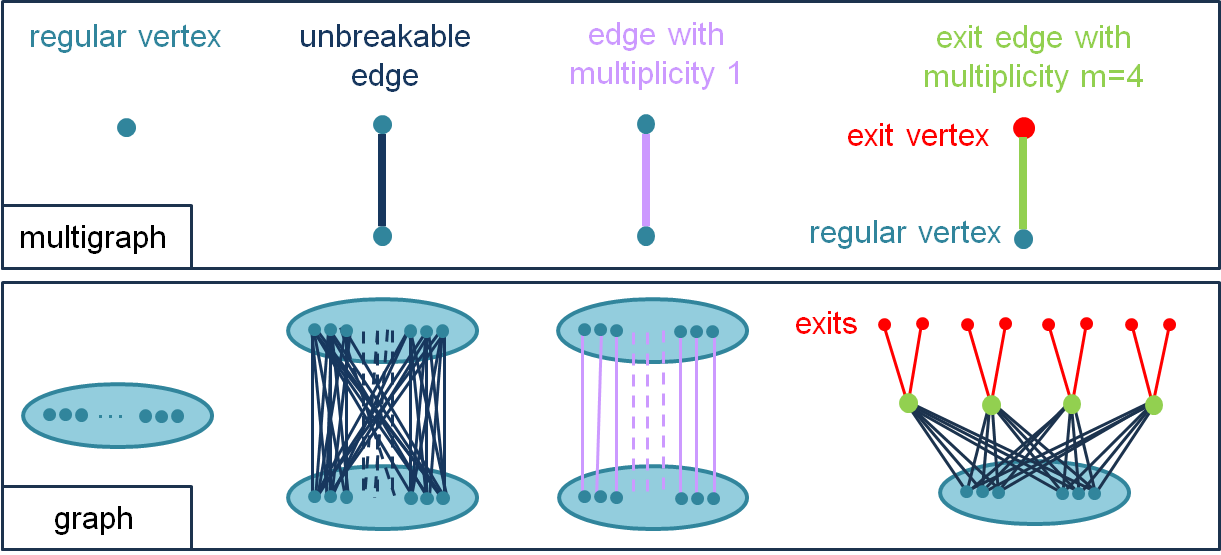}
  }
	\end{center}
	\caption{\label{translation} \textbf{Gadgets for translating our multigraph instance of Nemesis into an equivalent Nemesis graph instance.} Each regular vertex of the multigraph is translated into $N$ copies of the vertex with larger than the initial number of vertices of the graph. Unbreakable edges are translated in bicliques. Edges of multiplicity $1$ are translated in matchings. Exit edges of multiplicity $k$ are encoded through $k$ new regular vertices (the exit can be disabled by removing one exit edge of each new regular vertex).  }
\end{figure}

We now explain how we translate each element of the reduced multigraph instance of Subsection~\ref{multi} in order to build an equivalent simple graph instance (the translation gadgets are illustrated in Fig.~\ref{translation}).

\begin{itemize}
    \item \textbf{Vertices.}
    Each vertex of the multigraph instance, except the exits, is replaced by $N$ copies.
    More precisely, a vertex $u$ in the multigraph is replaced by $N$ vertices $u_1,\ldots,u_N$.

    \item \textbf{Unbreakable edges.}
    An unbreakable edge connecting vertices $u$ and $v$ in the multigraph is replaced by a complete bipartite graph between the sets $\{u_i\}_{1 \leq i \leq N}$ and $\{v_j\}_{1 \leq j \leq N}$.
    This yields $N^2$ edges instead of a single multiedge of multiplicity $N$.

    With such a construction, if the fugitive stands at some vertex $u_i$, the Nemesis cannot prevent him from reaching a copy of $v$, as blocking all $N$ possibilities would require too many rounds.
    This simulates the unbreakable edges of the multigraph instance.

    \item \textbf{Edges of multiplicity $1$.}
    An edge of multiplicity $1$ between vertices $u$ and $v$ in the multigraph is replaced by a matching connecting each $u_i$ to the corresponding $v_i$.
    This produces exactly $N$ simple edges.

    The key point is that if the fugitive stands at a vertex $u_i$, the Nemesis can delete the edge $(u_i,v_i)$ in a single round.
    Although the fugitive could then attempt to reach another copy $v_{j}$ via $u_{j}$, the Nemesis can repeat the same action.
    Since the timing of the reduction is tight, such repeated attempts would inevitably cause the fugitive to lose.
    Consequently, once the edge $(u_i,v_i)$ is deleted while the fugitive stands at $u_i$, the fugitive must follow an alternative path (in the electric gadget, this corresponds to an unbreakable derivation path, now implemented via a complete bipartite graph).

    Note that the Nemesis cannot efficiently delete these edges in advance.
    If she deletes $(u_i,v_i)$ prematurely, the fugitive may simply use another pair $(u_{j},v_{j})$.
    This limitation does not affect the correctness of the reduction.
    Indeed, in the multigraph construction, the winning strategy of the Nemesis deletes edges of multiplicity $1$ only when the fugitive stands at one of their endpoints.
    Therefore, this simulates edges of multiplicity $1$ from the multigraph instance.

    \item \textbf{Exit edges of arbitrary multiplicity.}
    To replace an exit edge of multiplicity $k$ between a vertex $u$ and an exit $t$, we introduce an \emph{exit gadget}.
    We need to take into account the fact that the vertex $u$ of the multigraph is represented in the graph by $N$ copies $u_i$.
    We additionally introduce $k$ intermediate vertices $p_j$, for $1 \leq j \leq k$.
    Each vertex $p_j$ is connected to exactly two exits.

    This gadget simulates an exit edge of multiplicity $k$ as follows.
    If the Nemesis has spent $k$ rounds deleting exit edges, she can deactivate the gadget by removing one exit edge incident with each $p_j$.
    In this case, whenever the fugitive reaches some $p_j$, the remaining exit edge can be deleted immediately, preventing escape.

    If, on the other hand, the Nemesis has used fewer than $k$ rounds on this gadget, then at least one vertex $p_j$ still has both exit edges intact.
    When the fugitive stands at some $u_i$, he can move to such a vertex $p_j$ and escape before the Nemesis can react.

    The Nemesis might attempt to delete edges between the vertices $u_i$ and the $p_j$ in advance.
    However, since the fugitive can choose among the $N$ copies of $u$, he can always reach a copy from which all remaining $p_j$ are still accessible.
    Hence, the Nemesis has no incentive to attack these edges.

    Overall, this exit gadget simulates an exit edge of multiplicity $k$.
\end{itemize}

The reduction of a \textsc{QSAT} instance is performed in two steps.
First, the \textsc{QSAT} instance is reduced to an equivalent instance of \textsc{Nemesis} on a multigraph.
Then, this multigraph instance is translated into an equivalent instance of \textsc{Nemesis} on a simple graph.

This translation is achieved by replacing each vertex and each multiedge of the multigraph by the corresponding vertex and edge gadget described above.
Each gadget simulates the behavior of its multigraph counterpart.

\begin{lemma}
The translation from the multigraph instance encoding an initial \textsc{QSAT} instance to the graph instance preserves the existence of a winning strategy for the fugitive or for the Nemesis.
\end{lemma}

As a consequence, the fugitive has a winning strategy in the multigraph instance if and only if he has a winning strategy in the resulting graph instance.
Since the size of the constructed graph is polynomial in the size of the initial \textsc{QSAT} instance, the reduction runs in polynomial time.
Together with the PSPACE-hardness of \textsc{Nemesis} on multigraphs and the fact that \textsc{Nemesis} belongs to PSPACE, this proves that \textsc{Nemesis} is PSPACE-complete on graphs (Theorem~\ref{t3}).

\subsection{PSPACE-completeness of the Cat Herding Problem}\label{pspace3}

The PSPACE-completeness of \textsc{Cat Herding} follows as a corollary of the PSPACE-completeness of \textsc{Nemesis} with two exits.
In the \textsc{Cat Herding} problem, the fugitive is a cat and the Nemesis is the cat herder.
Unlike \textsc{Nemesis}, the underlying graph has no exits.
The goal of the cat herder is to trap the cat on a single vertex in as few rounds as possible, while the goal of the cat is to remain free for as many rounds as possible.

\begin{figure}[ht]
  \begin{center}{
  \includegraphics[width=0.60\textwidth]{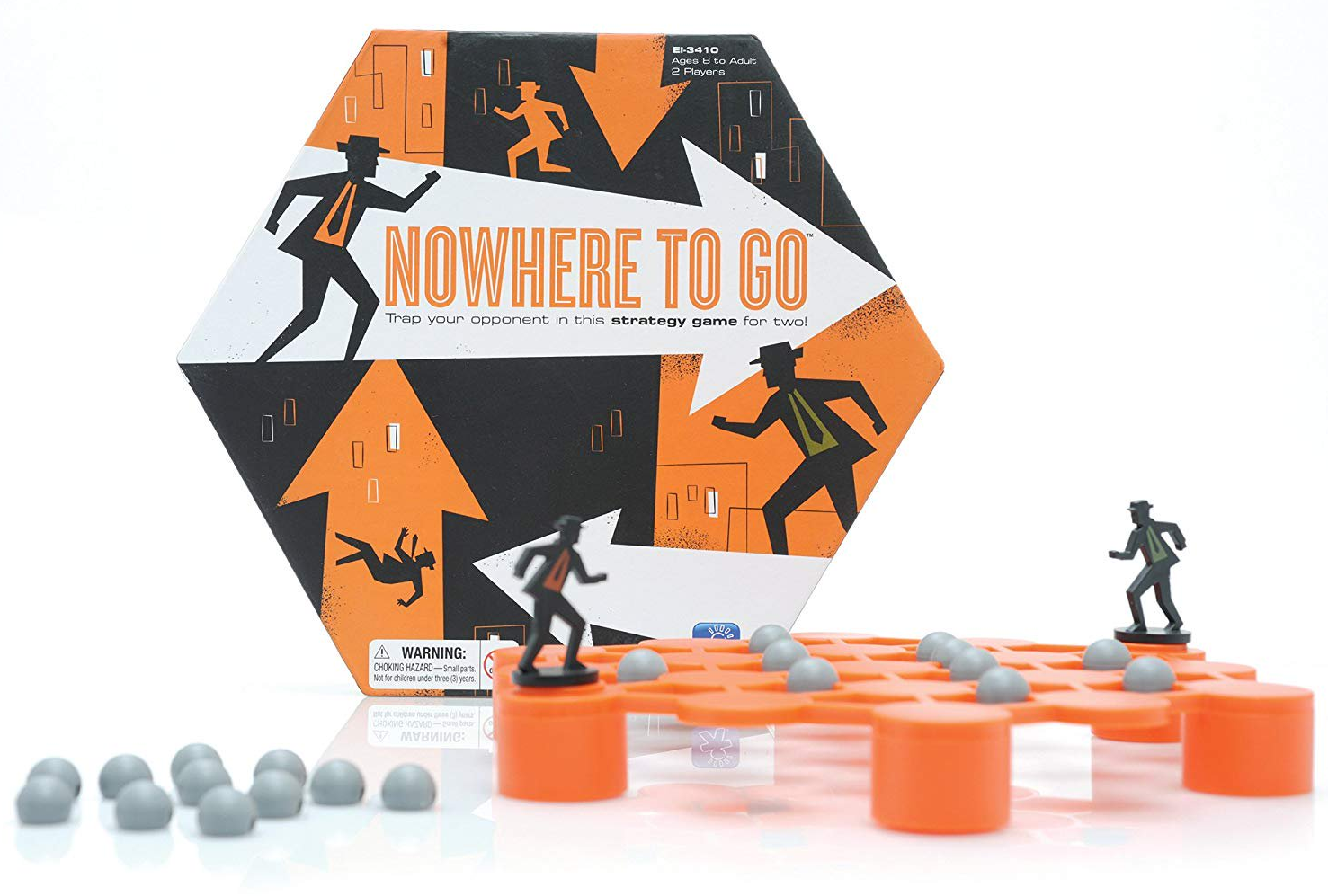}}
	\end{center}
	\caption{\label{cartherder} \textbf{The \textit{Nowhere To Go} board game~\cite{nowhereToGo}} (a game created by Hank Atkins and edited in 2012) which seems to have inspired the \textsc{Cat Herding} problem \cite{catphd}.}
\end{figure}

To establish PSPACE-completeness, we consider the associated decision problem: given an integer $k$, does the cat have a strategy to remain free for more than $k$ rounds?
We reduce \textsc{Nemesis} to \textsc{Cat Herding}.
An instance of \textsc{Nemesis} is given by a graph $G$ with $m$ edges,
a set of exits $E$, and a starting vertex $s$.
By Theorem~\ref{twoexits}, we may assume without loss of generality that $|E|=2$.
Let $\mathrm{cat}_s(G)$ denote the maximum number of rounds the cat can remain free in $G$
when starting from $s$.
Clearly, $\mathrm{cat}_s(G) \le m$, since the cat herder deletes one edge per round.

We now construct an instance $G'$ of \textsc{Cat Herding} from $G$.
The graph $G'$ is obtained by attaching a clique $K_{2m}$ to each exit of $G$,
identifying the exit vertex with one vertex of the corresponding clique.
The size of each clique is polynomial in $m$ and sufficiently large to ensure that,
once the cat enters a clique, it can remain free for strictly more than $2m$ rounds,
even if the cat herder has previously deleted up to $m$ edges of the clique.

If the fugitive has a winning strategy in $G$, then the cat can reach one of the two cliques in $G'$.
Once inside a clique, the cat can remain free for more than $2m$ rounds, and thus
$\mathrm{cat}_s(G') > 2m .$

Conversely, if the Nemesis has a winning strategy in $G$, then the cat herder can prevent the cat
from reaching either clique.
This takes at most $m$ rounds.
Once the cat is confined to $G$, the cat herder can delete the remaining edges of $G$
and trap the cat in at most $m$ additional rounds.
Therefore,
$\mathrm{cat}_s(G') \le 2m .
$

It follows that the answer to the decision problem asking whether the cat can remain free
for more than $2m$ rounds in $G'$ is positive if and only if the fugitive has a winning strategy
in the original \textsc{Nemesis} instance.
This establishes a polynomial-time reduction from \textsc{Nemesis} to \textsc{Cat Herding}.

Finally, \textsc{Cat Herding} is a finite two-player game with perfect information
and polynomially bounded play length.
Hence, determining the winner can be done in polynomial space.
Since \textsc{Nemesis} is PSPACE-complete (Theorem~\ref{t4}),
we conclude that \textsc{Cat Herding} is PSPACE-complete.

\section*{Perspectives}

We introduced \textsc{Nemesis} and \textsc{Blizzard}, with the somewhat surprising observation that these very simple games had not been previously investigated. We established several complexity results for different classes of graphs, but many open questions remain.
Can the PSPACE-completeness of \textsc{Nemesis} be established for planar simple graphs? Do polynomial-time algorithms exist for planar graphs? Or for planar graphs with a fixed number of exits?
The key feature of  \textsc{Nemesis} and \textsc{Blizzard} is their simplicity, which suggests that understanding their structure may help shed light on other problems, as illustrated by the complexity result about the \textsc{Cat Herding} problem.
Finally, we hope that these results illustrate how deceptively simple games on graphs can provide a rich playground for complexity theory while remaining fun to play.

\bibliography{biblio}

\end{document}